\documentclass[twocolumn]{aastex631}

\usepackage [english]{babel}
\usepackage [autostyle, english = american]{csquotes}
\MakeOuterQuote{"}

\usepackage{amsmath}

\begin{document}

\title{The cold Jupiter eccentricity distribution is consistent with EKL driven by stellar companions}

\correspondingauthor{Grant C. Weldon}
\email{gweldon@astro.ucla.edu}

\author{Grant C. Weldon}
\affiliation{Department of Physics and Astronomy, UCLA, Los Angeles, CA 90095, USA}
\affiliation{Mani L. Bhaumik Institute for Theoretical Physics, Department of Physics and Astronomy, UCLA, Los Angeles, CA 90095, USA}

\author{Smadar Naoz}
\affiliation{Department of Physics and Astronomy, UCLA, Los Angeles, CA 90095, USA}
\affiliation{Mani L. Bhaumik Institute for Theoretical Physics, Department of Physics and Astronomy, UCLA, Los Angeles, CA 90095, USA}

\author{Bradley M. S. Hansen}
\affiliation{Department of Physics and Astronomy, UCLA, Los Angeles, CA 90095, USA}
\affiliation{Mani L. Bhaumik Institute for Theoretical Physics, Department of Physics and Astronomy, UCLA, Los Angeles, CA 90095, USA}

\begin{abstract}

The large eccentricities of cold Jupiters and the existence of hot Jupiters have long challenged theories of planet formation. A proposed solution to both of these puzzles is high-eccentricity migration, in which an initially cold Jupiter is excited to high eccentricities before being tidally circularized. Secular perturbations from an inclined stellar companion are a potential source of eccentricity oscillations, a phenomenon known as the Eccentric Kozai-Lidov (EKL) mechanism. Previous studies have found that the cold Jupiter eccentricity distribution produced by EKL is inconsistent with observations. However, these studies assumed all planets start on circular orbits. Here, we revisit this question, considering that an initial period of planet-planet scattering on $\sim$Myr timescales likely places planets on slightly eccentric orbits before being modulated by EKL on $\sim$Myr-Gyr timescales. Small initial eccentricities can have a dramatic effect by enabling EKL to act at lower inclinations. We numerically integrate the secular hierarchical three-body equations of motion, including general relativity and tides, for populations of cold giant planets in stellar binaries with varied initial eccentricity distributions. For populations with modest initial mean eccentricities, the simulated eccentricity distribution produced by EKL is statistically consistent with the observed eccentricities of cold single-planet systems. The lower eccentricities in a multi-planet control sample suggest planetary companions quench stellar EKL. We show that scattering alone is unlikely to reproduce the present-day eccentricity distribution. We also calculate predictions for the inclinations and stellar obliquities in binary systems with cold Jupiters.

\end{abstract}

\section{Introduction} \label{sec:intro}

\begin{figure}[t!]
\begin{center}

\includegraphics[width=3.3in]{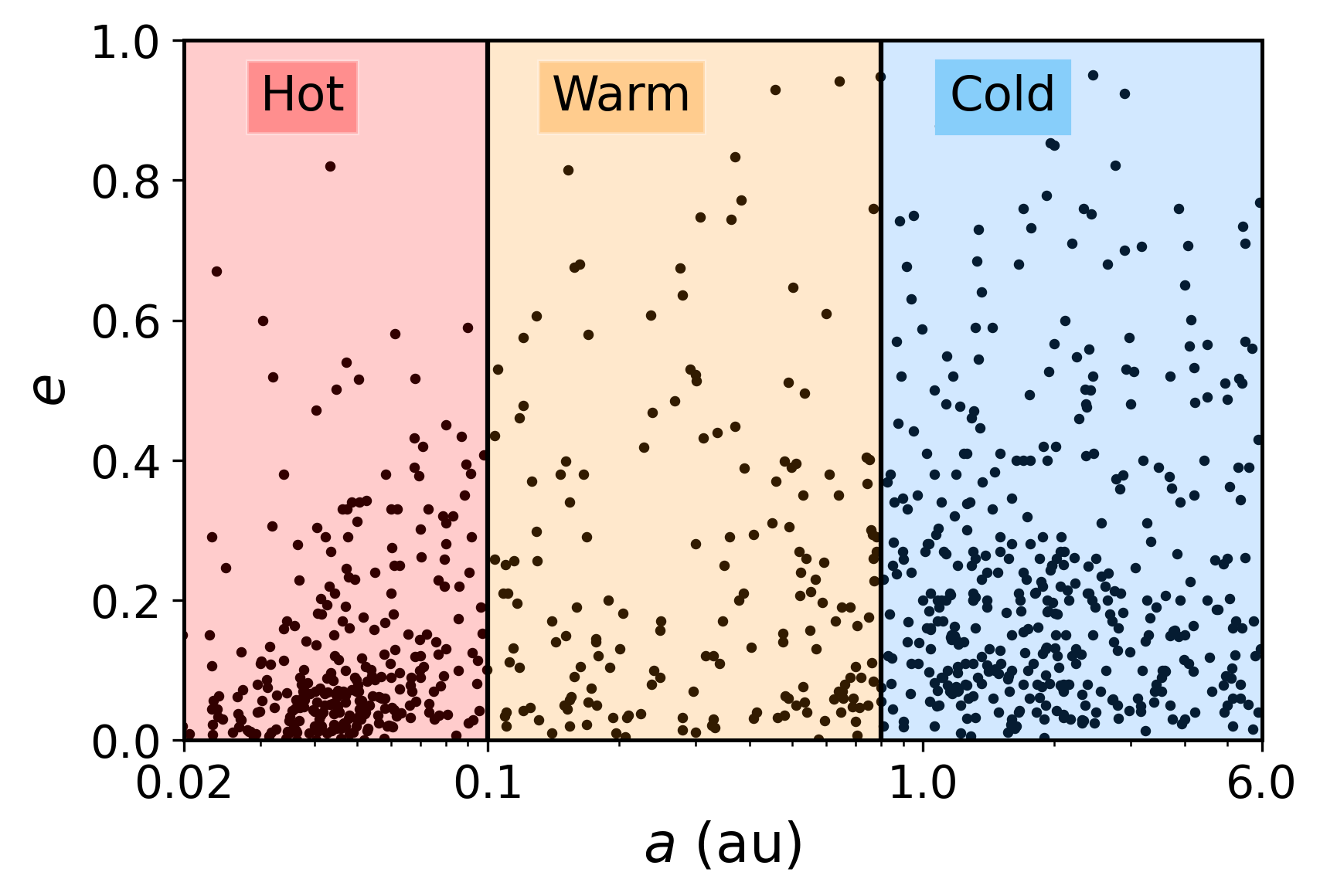}
\caption{\footnotesize The eccentricities and semi-major axes of observed planets (black dots) with mass $0.3-10 M_J$. In this work, we consider hot Jupiters to have $a<0.1$ au (red region), warm Jupiters to have $0.1$ au $<a<$ 0.8 au (orange region), and cold Jupiters to have $0.8$ au $<a<$ 6 au (blue region).} 
\label{fig:observed_jupiters}
\end{center}
\end{figure}

Thousands of exoplanets have been discovered in recent decades, with many exhibiting properties unlike the planets in our solar system. The presence of many giant planets on short-period orbits (so-called "hot Jupiters") has been particularly challenging to explain due to the hostile conditions for giant planet formation near stars \citep[for a review, see][]{Dawson+18}. Furthermore, giant planets at greater distances ("cold Jupiters") have been found to have a large range of eccentricities, unlike the nearly circular orbits of the solar system giants \citep[for a review, see][]{Winn+15}. A relative dearth of "warm Jupiters" with intermediate periods has also been found \citep[e.g.,][]{Jones+03,Udry+03}. In Fig. \ref{fig:observed_jupiters}, we show the eccentricities and semi-major axes of the observed Jupiter population \citep[from the NASA Exoplanet Archive, e.g.,][]{Akeson+13, PSCompPars}. We divide the population into hot Jupiters ($<0.1$~au), warm Jupiters ($0.1$ to $0.8$ au), and cold Jupiters ($0.8$ to $6$ au)\footnote{The warm/cold boundary chosen here is motivated by the location at which general relativity tends to quench secular perturbations, as discussed more in Appendix \ref{app:WC_boundary}.}.

High-eccentricity tidal migration has been proposed to explain the observed features of the Jupiter population \citep[e.g.,][]{Rasio+96,Holman+97,Wu+03,Fabrycky+07}. In this formation channel, an initially cold Jupiter is born at a large distance from its host star. Dynamical eccentricity excitations then cause the planet to undergo close pericenter passages, allowing tides to shrink and circularize the orbit until the planet arrives as a close-in hot Jupiter. A number of dynamical mechanisms have been proposed as the source of the eccentricity excitations needed to form hot Jupiters, including planet-planet scattering \citep[e.g.,][]{Rasio+96,Ford+05,Zhou+07,Juric+08,Chatterjee+08,Nagasawa+08,Ford+08,Nagasawa+11,Carrera+19}, secular chaos \citep[e.g.,][]{Wu+11,Teyssandier+19}, and secular perturbations from a faraway companion \citep[e.g.,][]{Holman+97,Wu+03,Takeda+05,Fabrycky+07,Naoz+11,Naoz+12,Lithwick+11,Katz+11,Teyssandier+13,Petrovich+15a,Petrovich15b,Anderson+16,Stephan+18,Stephan+20,Klein+24,Weldon+24,Stephan21,Stephan+24}. 

The latter phenomenon is known as the Eccentric Kozai-Lidov (EKL) mechanism \citep[e.g.,][]{Kozai62,Lidov62,Naoz16}. In this mechanism the faraway companion of either planetary or stellar nature secularly excites eccentricity and inclination oscillations. Many of the aforementioned studies focused on the potential for EKL in stellar binaries to form hot Jupiters. Indeed, Hot Jupiters have been found to preferentially exist in stellar binaries \citep[e.g.,][]{Knutson+14,Ngo+15}, but it was argued that, in part due to the large population of hot Jupiters in single-star systems, only $\sim20$\% of Hot Jupiters could be formed through stellar EKL \citep[][]{Ngo+15}. 
However, recent work by \cite{Stephan+24} suggested that the majority of seemingly single hot Jupiters could have formed through EKL in binaries that were later unbound by white dwarf kicks.

In addition to forming hot Jupiters, EKL should also leave its signature on the cold Jupiter population. \cite{Takeda+05} and \cite{Petrovich15b} calculated the eccentricity distribution of cold Jupiters shaped by stellar EKL. In both studies, EKL produced an overabundance of low eccentricity planets and an underabundance of moderate to high eccentricity planets. It was again concluded in these works that only a small fraction of Hot Jupiters could be formed through EKL in stellar binaries.

Here, we show that stellar EKL plays a more dominant role in sculpting the cold Jupiter eccentricity distribution than previously claimed. While this question was briefly addressed in the literature by \cite{Takeda+05} and \cite{Petrovich15b}, we revisit this question, utilizing two key differences in our approach. 
\begin{itemize}
    \item First, we allow for an initial Rayleigh distribution of cold Jupiter eccentricities. Previous studies assumed all planets start on initially circular orbits. In reality, an early period of planet-planet scattering likely places many planets on eccentric orbits. The typical outcome of scattering is a Rayleigh distribution, and the mean eccentricity depends on factors such as planet multiplicity and mass ratios \citep[e.g.,][]{Rasio+96,Adams+03,Zhou+07,Ford+08,Chatterjee+08,Nagasawa+08,Malmberg+09,Raymond+10}. While these works attempt to reproduce the present-day eccentricity distribution through scattering alone, it is physically likely that planets undergo a relatively quick period of scattering (timescales $\sim10^5 - 10^7$ yr) before being modulated by EKL on characteristically longer timescales (up to $\sim10^9 - 10^{10}$ yr). In this work, we show that small initial eccentricities can alter the resultant distribution predicted by EKL and produce one that is statistically consistent with the observed population.
    
    \item We consider the eccentricities of all cold planets with semi-major axes $>0.8$ au, regardless of the final outcome of the system. We include proportional contributions to the eccentricity distribution of non-migrating cold planets, cold planets that eventually undergo migration to become hot Jupiters, and cold planets that are ultimately destroyed by tidal disruption.

\end{itemize}

The details of the numerical simulations, initial conditions, and observed sample are discussed in \S \ref{sec:setup}. We compare the results of the numerical simulations with the observations in \S \ref{sec:results}. Comparisons with other studies and implications are discussed in \S \ref{sec:discussion}. We summarize our work and conclusions in \S \ref{sec:conclusion}.

\section{Numerical setup} \label{sec:setup}

\subsection{Numerical methods}
\label{subsec:methods}

We numerically solve the octupole-level secular equations for the hierarchical three-body system following \citet{Naoz+11,Naoz+13}. A hierarchical system has a relatively tight ``inner'' binary (stellar mass $m_1$ and planet mass $m_2$) with semi-major axis $a_1$ orbited by a faraway ``outer'' companion (in this work, a star) with mass $m_3$ and semi-major axis $a_2$. We denote the angle of inclination of the inner (outer) orbit with respect to the total angular momentum by $i_1$ ($i_2$), and the mutual inclination between the two orbits is $i$ = $i_1 + i_2$. 

For hierarchical systems, the three-body Hamiltonian can be averaged over the orbital periods and expanded in powers of the small semi-major axis ratio $\alpha = a_1/a_2$ \citep[e.g.,][]{Kozai62, Harrington68, Ford+00, Naoz+13}. At the quadrupole level (proportional to $\alpha^2$), the inner orbit can undergo oscillations of eccentricity and inclination on timescales much longer than the orbital periods \citep{Kozai62, Lidov62}. The quadrupole level of approximation is often insufficient when the outer companion's orbit is eccentric or when the planet mass is non-negligible \citep{Naoz+13}. In these cases, the octupole contribution (proportional to  $\alpha^3$) can drive the inner orbit to extremely high eccentricities or even flip from prograde to retrograde with respect to the total angular momentum \citep[see for a full set of equations,][]{Naoz16}. 

In addition to EKL, we include general relativistic precession of the inner and outer orbit \citep[][]{Naoz+13b}. Equilibrium tides are also included following \cite{Eggleton98} and \cite{Eggleton+01}. We fix the viscous timescale $t_{v,1} = 50$ years for the star and $t_{v,2} = 0.01$ years for the planet \citep[e.g.,][]{Petrovich15b}. We define the stellar obliquity $\psi$ as the angle between the spin of the inner star and the direction of the angular momentum of the inner orbit. The sky-projected obliquity $\lambda$ can be measured through the Rossiter-McLaughlin effect \citep[e.g.,][]{Gaudi+07}. Our code follows the precession of the spin vector \citep[again see for full set of equations,][]{Naoz16}. Furthermore, we model the stellar evolution of the stars using the SSE evolution code of \cite{Hurley2000}. The combined code with secular evolution, general relativity, tides, and stellar evolution has been tested and applied to various astrophysical systems \citep[e.g.,][]{Naoz16,Naoz+16,Stephan+16,Stephan+18,Stephan+19,Stephan+20,Stephan21,Angelo+22,Shariat+23,Shariat+24}. We show an example time evolution for an individual system in Appendix \ref{app:time_evolution}.

The upper limit for each system's integration time in our simulations is $10$ Gyr. A planet is tidally disrupted if it crosses the Roche limit \begin{equation} \label{eq:roche}
    R_{\rm Roche} = \eta_{\rm Roche} R_J \left(\frac{m_1+m_2}{m_2}\right)^{1/3} \ ,
\end{equation}
where $\eta_{\rm Roche} = 2.7$ \citep[][]{Guillochon+11}. If $a_1 (1-e_1) < R_{\rm Roche}$, we stop the integration and consider the planet lost. We also consider the planet lost if it is engulfed during stellar evolution. If a planet is lost, we still include the planet's proportional contribution to the eccentricity distribution prior to being lost in our simulated sample, as it is possible that some cold Jupiters are observed en route to being destroyed. If a hot Jupiter forms, the orbit circularizes until the integration is stopped once $1-e_1 < 10^{-4}$, and we consider the hot Jupiter to remain in place for the remaining time steps. We note that for the purposes of the cold Jupiter eccentricity distribution, the planet "leaves" the distribution once the semi-major axis shrinks below 0.8 au. 

We sample the output of all integrations equally in time. However, to ensure sufficient resolution for all integrations to proceed, a different underlying time step is used for each system. We first compute the quadrupole timescale, corresponding to the period of Kozai-Lidov oscillations \citep[e.g.,][]{Antognini15}
\begin{equation}
    t_{\mathrm{quad}} = \frac{16}{15} \frac{a_2^3\left(1-e_2^2\right)^{3 / 2} \sqrt{m_1+m_2}}{a_1^{3 / 2} m_3 k} \ ,
    \label{eq:t_quad}
\end{equation} 
where Newton's constant is $k^2$. We then use a time step of $t_{\rm quad}/100$, rounded in years to the nearest power of ten. For example, if $t_{\rm quad}/100 = 3 \times 10^4$ years, we use a time step of $10^4$ years for that system. If $t_{\rm quad}/100$ exceeds $10^5$ years, we impose a time step of $10^5$ years. For all systems, including those with shorter underlying time steps, we print out the integration every $10^5$ years to ensure that the final distributions are evenly and equally sampled in time. This value is chosen to prevent the output from becoming computationally expensive, while also providing enough resolution for the vast majority of systems that are tidally disrupted on short timescales.

\begin{table*}
\begin{center}
\setlength{\tabcolsep}{7pt}
\caption{Summary of Simulations}
\label{tab:sims}
\begin{tabular}{c c c c c}
\hline\hline
Name &  $e_1$ (initial) & $a_1$ (initial) [au] & $N_{\rm sims}$ & KS $p$-value $^b$ \\
\hline

circ & 0.01 & Uniform(0.5,6) & 250 & $10^{-6}$\\
r0.05 & Rayleigh ($\mu$ = 0.05) & Uniform(0.5,6) & 200 & $10^{-6}$ \\
r0.09 & Rayleigh ($\mu$ = 0.09) & Uniform(0.5,6) & 500 & 0.02 \\
r0.13 & Rayleigh ($\mu$ = 0.13) & Uniform(0.5,6) & 800 & 0.50 \\
5AUr0.13 & Rayleigh ($\mu$ = 0.13) & 5 & 200 & 0.66 \\
r0.175 & Rayleigh ($\mu$ = 0.175) $^{a}$ & Uniform(0.5,6) & 250 & 0.49 \\
r0.21 & Rayleigh ($\mu$ = 0.21) & Uniform(0.5,6) & 250 & 0.01 \\
r0.25 & Rayleigh ($\mu$ = 0.25) & Uniform(0.5,6) & 200 & $10^{-3}$ \\
beta & Beta ($\alpha=1$, $\beta = 6$, $\mu \sim$ 0.15) & Uniform(0.5,6) & 200 & 0.51 \\
2state & 50\% 0.01, 50\% Rayleigh ($\mu$ = 0.25) & Uniform(0.5,6) & 200 & 0.04 \\

\hline \hline
\end{tabular}
\end{center}
$^a$ This value is motivated by \cite{Moorhead+11}. \\
$^b$ The respective simulated eccentricity distribution is compared with the MSSP observed sample.
\end{table*}

\subsection{Initial conditions}
\label{subsec:inicons}

To explore the eccentricity distribution for populations of cold planets subject to EKL, we run 10 sets of numerical simulations, which are described in Table \ref{tab:sims}. For the initial conditions, we draw the stellar masses $m_1$ and $m_3$ between 0.6 $M_{\odot}$ and 1.6 $M_{\odot}$ (roughly corresponding to FGK stars) from a Salpeter distribution with $dn/dm \propto m^{-2.35}$ \citep{Salpeter55}. We draw the planet mass $m_2$ between 0.3$M_J$ and 10$M_J$ from a power-law distributrion with $dn/dm \propto m^{-1}$ \citep{Marcy+2000}. The stellar radius is set by the SSE code, and we fix the initial radius of the planet to be $1 R_J$. We draw $a_1$ uniformly from $0.5-6$ au in all simulations, except one in which we explore the effect of initially setting $a_1 = 5$ au. We draw outer orbit periods from the log-normal distribution of \cite{Duquennoy+91} and require that $a_2$ be between 50 and 1500 au. The outer orbit eccentricity $e_2$ is drawn uniformly from 0 to 1 \citep[e.g.,][]{Raghavan+10,Moe+17}\footnote{Other works have modeled the binary eccentricity distribution as thermal \citep[e.g.,][]{Jeans19}. The thermal distribution is more heavily weighted towards higher eccentricities and consequently leads to even stronger EKL effects. We take a more conservative approach by adopting a uniform eccentricity distribution.}, and the initial mutual inclination is sampled isotropically (uniform in $\cos i$ from $0-180^{\circ}$). The initial arguments of periapse $\omega_1$ ($\omega_2$) of the inner (outer) orbit are sampled uniformly from $0-360^{\circ}$. In all of our simulations, the stellar spin and planet orbit are initially aligned, such that $\psi$ = 0.001$^{\circ}$.

We explore the effect of varying the initial distribution of the planet's orbital eccentricity $e_1$. A first set of simulations is started on nearly circular orbits ($e_1 = 0.01$ for all systems), similar to \cite{Takeda+05} and \cite{Petrovich15b}. In other sets of simulations, we draw the initial eccentricity from Rayleigh distributions with means varying from 0.05 to 0.25. Under the assumption of planet-planet scattering dominated by chaotic diffusion, a Rayleigh distribution is the analytically expected outcome \citep[][]{Zhou+07}. Indeed, a Rayleigh distribution is produced in numerical simulations of scattering \citep[e.g.,][]{Zhou+07,Juric+08,Chatterjee+08}. To test a distribution with a different shape, we also perform one set of simulations with $e_1$ drawn from a Beta distribution with parameters $\alpha=1$ and $\beta=6$. This choice was motivated by the fact that exoplanet eccentricities resemble a Beta distribution \citep{Kipping13}, but was arbitrarily constructed to have a small, but non-negligible mean of $\sim0.15$. We also test a two-state model, in which 50\% of systems begin with $e_1 = 0.01$ and 50\% of systems have $e_1$ drawn from a Rayleigh distribution with mean 0.25. While these values are again selected somewhat arbitrarily, here, we are motivated by the fact that there may be distinct dynamically active and inactive populations of scattered planets \citep[e.g.,][]{Juric+08}.

We note that all of our initial conditions are stable according to the \cite{Mardling+01} stability criterion
\begin{equation}
    \frac{a_2}{a_1}>2.8\left(1+\frac{m_3}{m_1+m_2}\right)^{2 / 5} \frac{\left(1+e_2\right)^{2 / 5}}{\left(1-e_2\right)^{6 / 5}}\left(1-\frac{0.3 i}{180^{\circ}}\right) \ .
\end{equation}
The systems are also sufficiently hierarchical. That is, the hierarchy parameter $\epsilon$ fulfills
\begin{equation}
\label{eq:eps}
    \epsilon = \frac{a_1}{a_2} \frac{e_2}{1-e_2^2} < 0.1 \ .
\end{equation}

\subsection{Observed sample}

To construct the observed sample, we use the NASA Exoplanet Archive \citep[e.g.,][]{Akeson+13, PSCompPars}. Our sample contains observed giant planets with mass $0.3-10$$M_J$ and with semi-major axis $0.8-6$ au that have measured eccentricities (i.e. error bars are reported for the measurement)\footnote{As some systems have multiple recorded measurements, we use the Composite Data Table, which effectively selects a default value for each system.}. Measured eccentricities often undergo significant revision following initial measurements. Therefore, we only consider planets discovered in 2020 or before to avoid uncertain initial eccentricity estimates.

We next split the sample into subsets, each with a different cut on planetary or stellar multiplicity. We consider the following cases: multi-star, single-planet (MSSP, 58 planets), single-star, single-planet (SSSP, 200 planets), and multi-planet (MP, 94 planets, regardless of stellar multiplicity). To make a faithful comparison to the hierarchical triple simulations, MSSP is taken to be the default sample. For our purposes, we consider a single-planet system as having one giant planet beyond the cold threshold of 0.8 au, even if there are planets closer in, as widely separated planets respond to EKL independently. For the MP sample, we aim to construct a control sample of giant planets in which mutual gravitational interactions are significant compared to potential stellar EKL. As such, from the MP sample we remove systems in which the ratio of semi-major axes between the outer two planets is $>5$, leaving a sample of compact multi-planet systems. We manually ensure that planets with the same host stars under different labels, as well as planets that orbit separate stars within the same binary system, are appropriately accounted for. We note that some systems may have undetected planetary or stellar companions. The effects of planetary and stellar multiplicity are explored in \S \ref{subsec:multiplicity}.

Nearly every system in the sample was discovered with the radial velocity method.  While high-eccentricity orbits are more difficult for radial velocity surveys to sample in time, simulations of detection thresholds have revealed that this effect is nearly canceled by the fact that they have higher radial velocity amplitudes \citep[][E. Petigura 2024, private communication]{Fischer+92,Shen+08}. As such, the observed eccentricity distribution is largely unbiased by eccentricity selection effects.

\section{Results} \label{sec:results}
\subsection{Eccentricity distribution}
\label{subsec:e_distribution}

\begin{figure*}[ht]
\begin{center}

\includegraphics[width=7.1in]{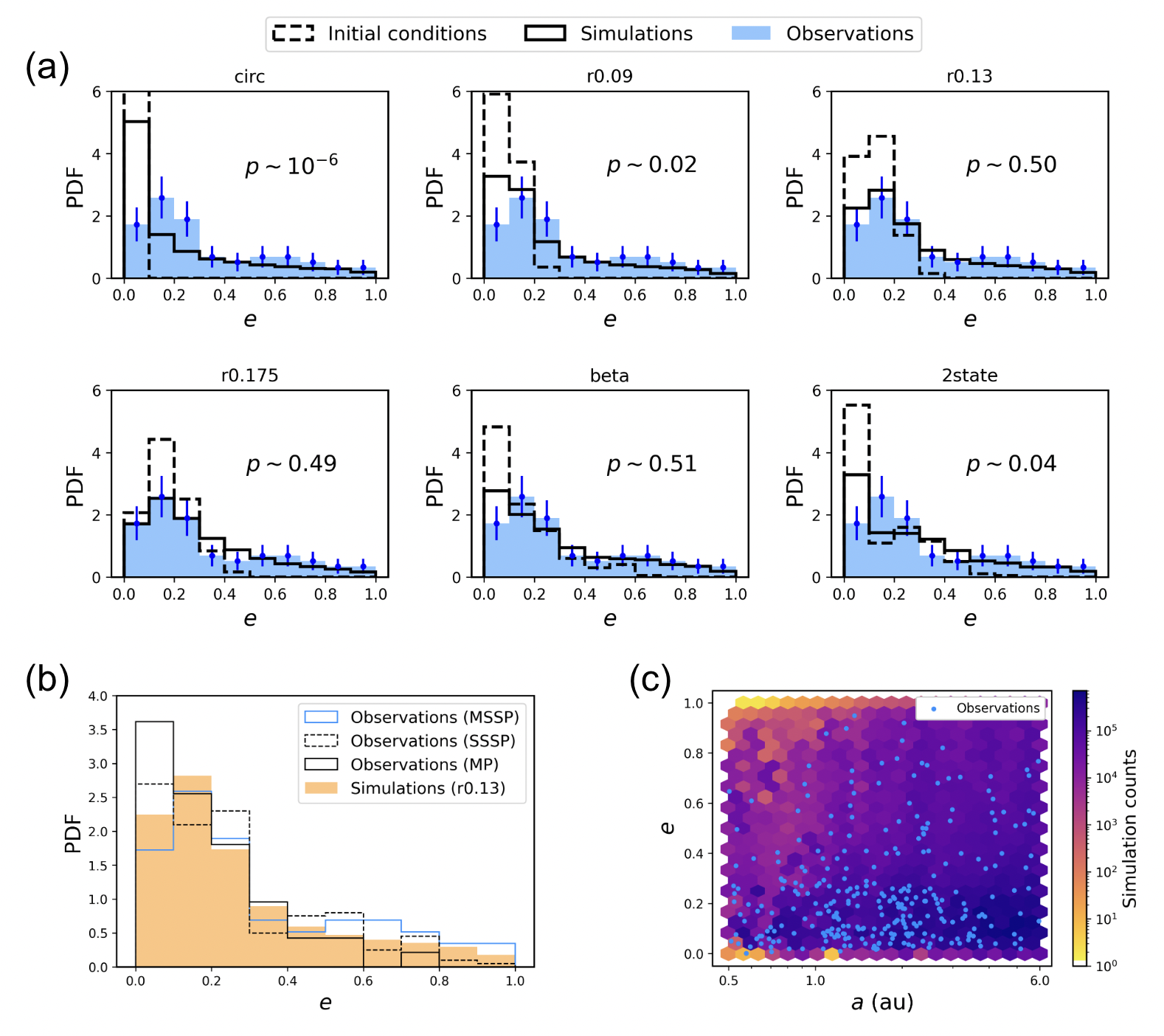}
\caption{\footnotesize Panel (a): Distributions of initial conditions (dashed black line), simulated eccentricities (solid black line), and observed eccentricities in the MSSP sample (shaded blue) for six different simulation runs. The Poisson uncertainty for each bin is shown in blue. Each plot is labeled with the name of the run. The KS-test $p$-values comparing the simulated and observed distributions are shown. Panel (b): Comparison of the simulated eccentricity distribution (shaded orange) in the \texttt{r0.13} simulations with the distribution for multi-star, single-planet systems (MSSP, solid blue line), single-planet, single-star systems (SSSP, dashed black line), and multi-planet systems (MP, solid black line). Panel (c): Two-dimensional histogram with simulated systems in the \texttt{r0.13} run binned based on eccentricity and semi-major axis. The color code shows the number of counts in each bin. Darker regions correspond to a higher amount of time spent in that region. Here, we show systems with $a>0.5$ au to examine the boundary between the warm and cold populations. The observed population of $0.3-10$ $M_J$ planets (blue dots, single-planet systems) is overplotted.} 
\label{fig:ecc_distributions}
\end{center}
\end{figure*}

We compare the observed eccentricity distribution (multi-star, single-planet sample) to the distribution of all cold planetary eccentricities obtained in each respective simulation run. The Kolmogorov-Smirnov (KS) test $p$-values are reported in Table \ref{tab:sims}. In panel (a) of Fig. \ref{fig:ecc_distributions}, for a selected number of our simulation runs, we plot the initial conditions (dashed black line), simulated eccentricity distribution (solid black line), and observed distribution (shaded blue region) with the Poisson uncertainty for each bin (blue lines). We note that the Poisson errors for the simulations are very small due to the large number of data points. The KS-test $p$-values for these runs are also shown.

For the systems that begin with nearly circular orbits (simulation runs \texttt{circ} and \texttt{r0.05}), we find that EKL produces a $6\sigma$ excess of systems with low eccentricities ($e_1<0.1$) and $2\sigma$ deficit of systems with moderate eccentricities ($0.1<e_1<0.3$). The very low KS-test $p$-values show that the null hypothesis that the simulated and observed data originate from the same distribution can be rejected. As expected, this result is largely consistent with the works by \cite{Takeda+05} and \cite{Petrovich15b} that started their planets on nearly circular orbits and found that too many remain on such orbits. 

For systems with Rayleigh distributions with slightly higher eccentricities, the distributions match much more closely. The \texttt{r0.09} run becomes more consistent with the data, having a $p$-value of 0.02, although a discrepancy remains due to a $3\sigma$ overabundance of low eccentricity orbits. The \texttt{r0.13} run is very similar to the observed data, having a $p$-value of 0.50. The \texttt{r0.175} run, motivated by the mean value in \cite{Moorhead+11}, yields a similar $p$-value of 0.49. In these cases, the null hypothesis that the simulated and observed data originate from the same distribution cannot be rejected. The discrepancies between the observed and simulated data in each bin in Fig. \ref{fig:ecc_distributions} are not statistically significant, as they are all within $1-2\sigma$ of the Poisson uncertainty. 

While the aforementioned systems had $a_1$ drawn uniformly from 0.5 to 6 au, we also perform a run (\texttt{5AUr0.13}) starting the planets all at 5 au, as there are still large uncertainties in the exact formation distance of giant planets. Here, we find a $p$-value of 0.66, suggesting that the simulated eccentricity distribution remains consistent with the observations. While we do not expect all planets to begin at 5 au, this result shows that the eccentricity distribution is not highly sensitive to the observationally ill-constrained initial semi-major axis of the planet.

We find that our runs with even higher mean eccentricities (\texttt{r0.21} and \texttt{r0.25}) are largely inconsistent with the observations, having $p$-values $\leq 0.01$. As such, the mean eccentricity for the initial Rayleigh distribution must be $\sim 0.1-0.2$ to produce a match to the observations. Below a mean of $\sim0.1$, too many systems are on nearly circular orbits. Above a mean of $\sim0.2$, too many systems are on moderate to high eccentricity orbits.

We also explore the effect of using an initial eccentricity distribution other than the Rayleigh distribution. In the $\texttt{beta}$ run, we draw $e_1$ from a Beta distribution with $\alpha=1$ and $\beta=6$, such that the mean eccentricity is $\sim0.15$. In contrast with the Rayleigh distribution, which is sharply peaked near its mean, the Beta distribution has a higher spread. Comparatively more planets start on near-circular orbits, and more start on very eccentric orbits. Despite this difference, we find that the resultant eccentricity distribution is consistent with the observations, having a $p$-value of 0.51. This is a similar significance to the best-fitting Rayleigh distribution (mean of 0.13). The $\texttt{beta}$ run produces a worse match at low eccentricities than the $\texttt{r0.13}$ run, but a better match at high eccentricities. We additionally test a two-state model in the \texttt{2state} run, in which 50\% of systems are dynamically inactive (beginning with $e_1 = 0.01$) and 50\% are more active (a Rayleigh distribution with mean 0.25). The eccentricity distribution produced has a $p$-value of 0.04 and is marginally consistent with the observations. The presence of many initially circular orbits leads to too many planets remaining on circular orbits at the 3$\sigma$ level. While this two-state model was constructed arbitrarily as a proof of concept, one could likely go further to tune one that produces a better match to the observations. Together, these results suggest that the initial eccentricity distribution does not necessarily need to be a pure Rayleigh distribution for EKL to produce a statistically consistent distribution with the observations. 

To assess how the uncertainties on individual eccentricity measurements may affect our conclusions, we resample the MSSP data given the reported upper and lower error bars to generate new observed distributions. For 1000 different realizations of the data, we perform a KS-test with the simulations. When comparing to the \texttt{r0.13} and \texttt{r0.175} simulations, which nominally agree with the observations, 94.9\% and 95.3\% of the realizations respectively still have a statistically significant agreement, indicating that our findings are largely robust with the measurement uncertainties considered.

\subsection{Planetary and stellar multiplicity}
\label{subsec:multiplicity}

We next examine the effects of stellar and planetary multiplicity on the eccentricity distribution. In panel (b) of Fig. \ref{fig:ecc_distributions}, we show the \texttt{r0.13} simulation distribution, along with the observed sample with different multiplicity cuts. The MSSP sample is shown with a solid blue line, the SSSP sample is shown with a dashed black line, and the MP sample is shown with a solid black line.

We first compare our simulated eccentricity distribution to the observed samples. We use both the Kolmogorov-Smirnov (KS) test and the Anderson-Darling (AD) test. The latter is more sensitive to the wings of the distributions, where we visually observe the largest differences. Comparing the simulations with the MSSP, SSSP, and MP samples, we respectively obtain $p$-values of 0.50, 0.10, and 0.03 from the KS-test and $p$-values of 0.25, 0.10, and 0.001 from the AD-test. The EKL model is strongly consistent with the MSSP sample, is a weaker but still significant match to the SSSP sample, and is largely inconsistent with the MP sample according to both tests.

We also compare the observed samples to each other without including the model comparison. To test for the effect of stellar multiplicity, we compare the MSSP and SSSP samples and obtain $p$-values of 0.41 (KS-test) and 0.14 (AD-test). These samples are statistically consistent with one another according to both tests, although we note that the SSSP is shifted slightly to lower eccentricities. To test against the MP control sample, in which stellar perturbations should be insignificant compared to mutual gravitational interactions, we compare the combined MSSP and SSSP samples with the MP sample. This comparison yields $p$-values of 0.07 (KS-test) and 0.02 (AD-test). These samples are consistent according to the KS-test, but are inconsistent with one another according to the AD-test due to the overabundance of low eccentricity planets and underabundance of high eccentricity planets in the MP sample.

We expect the isolated planets in multi-star systems to most strongly respond to EKL, and indeed, the simulations best match the MSSP sample. The SSSP sample shows evidence for the action of EKL, although the distribution is slightly shifted to lower eccentricities. This is consistent with the notion that the SSSP sample actually contains some systems with undetected perturbing stellar companions. To estimate the fraction of systems that may have a stellar perturber, we mix a pure Rayleigh distribution with mean 0.13 and the simulated \texttt{r0.13} distribution to heuristically model the single-star and multi-star components. We vary the mixing ratio and perform a KS-test between the mixed simulations and the observed SSSP sample. We find that 48\% of systems having a stellar perturber produces agreement at the $2\sigma$ level. We repeat for a Rayleigh distribution with mean 0.175 and the \texttt{r0.175} run and find that 38\% of systems may have a stellar perturber. These values are not particularly surprising given the $\sim$50\% binary fraction of Sun-like stars \citep[e.g.,][]{Raghavan+10}. While slightly higher than may be expected given that some binaries have already been accounted for in the MSSP sample and the lower multiplicity for wider binaries, the eccentricities in some systems may also be driven by undetected planetary companions \citep[e.g.,][]{Naoz+11,Petrovich+15a} or from an unbound stellar perturber \citep[e.g.,][]{Stephan+24}. We also note that our estimates for the companion fraction are sensitive to the tail of the distribution, where differences are within the Poisson error. As such, small fluctuations in the number of highly eccentric planets may significantly affect this estimate. More stringent constraints on the presence of planetary and stellar companions in the SSSP sample with future observations would help to test this work more robustly.

The MP control sample most strongly disagrees with the simulations and with the observed single-planet samples, as it contains too many low eccentricity planets and too few high eccentricity planets. We expect that the presence of a planetary companion hinders EKL oscillations from any stellar perturber \citep[e.g.,][]{Pu+18,Denham+19,Wei+21,Faridani+22}, and indeed the discrepancy we observe suggests that stellar EKL is responsible for shaping the eccentricity distribution in the single-planet samples. We note that in some configurations the outer planet itself may induce EKL oscillations on the inner planet \citep[e.g.,][]{Naoz+11,Petrovich+15a}, though the exact nature of the competition between stellar and planetary EKL is outside the scope of this work. It has been noted before \citep[e.g.,][]{Wright+09} that multi-planet systems tend to have lower eccentricities than single-planet systems. This trend was previously attributed to dynamical stability, as high eccentricities in multi-planet systems may lead to collisions or ejections. Here, we suggest that the effect of planetary companions quenching the precession induced by stellar perturbers is also at play.

Altogether, the results for different planetary and stellar multiplicities are broadly consistent with the expected outcomes if stellar EKL indeed shapes the eccentricity distribution of the cold Jupiters.

\subsection{Eccentricity and semi-major axis map of planets}

We now look at the map of planets in both eccentricity and semi-major axis space. In panel (c) of Fig. \ref{fig:ecc_distributions}, we construct a two-dimensional histogram of the eccentricity and semi-major axis in the \texttt{r0.13} run. The color code shows the number of simulation counts in each hexagonal bin (with darker colors corresponding to a higher number of counts), and the observed population of Jupiter-like planets (in single-planet systems) is overplotted. Here, we look at planets with semi-major axes larger than 0.5 au to examine the boundary between the warm and cold populations.

Our systems were initially set with semi-major axes chosen uniformly from 0.5 to 6 au. We see that between $\sim 0.5-1$ au, there are fewer simulated systems at eccentricities above $\sim0.4$ (light purple and orange regions) compared to larger semi-major axes (dark purple regions). There is also a dearth of observed systems in this region, which is especially notable considering the preference for radial velocity surveys to find closer-in planets. We note that this feature in the simulations remains even when a linear binning scheme is used, as the differences in bin counts are much larger than the differences in bin size. As discussed in Appendix \ref{app:WC_boundary}, we chose 0.8 au to be the threshold between warm and cold planets, as it analytically corresponds to the median value at which general relativistic precession overwhelms EKL. The relative paucity of observed high-eccentricity planets near this threshold aligns with the simulations, suggesting that, indeed, stellar EKL is less efficient at exciting planets to high eccentricities in this region. 

We perform a two-dimensional KS-test between the simulations and observations in this regime, considering both eccentricity and semi-major axis. Here, we only look at planets in multi-star systems to provide the most robust comparison. Due to the preference for radial velocity surveys to find closer-in planets, we heuristically model the detection bias in our simulations by removing data points with a probability based on the semi-major axis. We consider fall-offs of $a^{-1/2}$ \citep[the dependence of radial velocity amplitude on semi-major axis, see, e.g.,][]{Cumming+04} and $a^{-1}$ (as additional effects such as survey duration prefer short-period planets). A full treatment of detection bias for the many radial velocity surveys in the NASA Exoplanet Archive is beyond the scope of this work. For 10000 simulated data points, we obtain $p$-values of 0.02 and 0.15 for each respective fall-off case. Given the uncertain initial locations of planets and the basic model for detection bias, these matches are encouraging. These results suggest that giant planets may initially be found at distances of $\sim0.5-6$ au following scattering.

\section{Discussion}
\label{sec:discussion}

\subsection{EKL with small initial eccentricities}

\begin{figure}[]
\begin{center}

\includegraphics[width=3.3in]{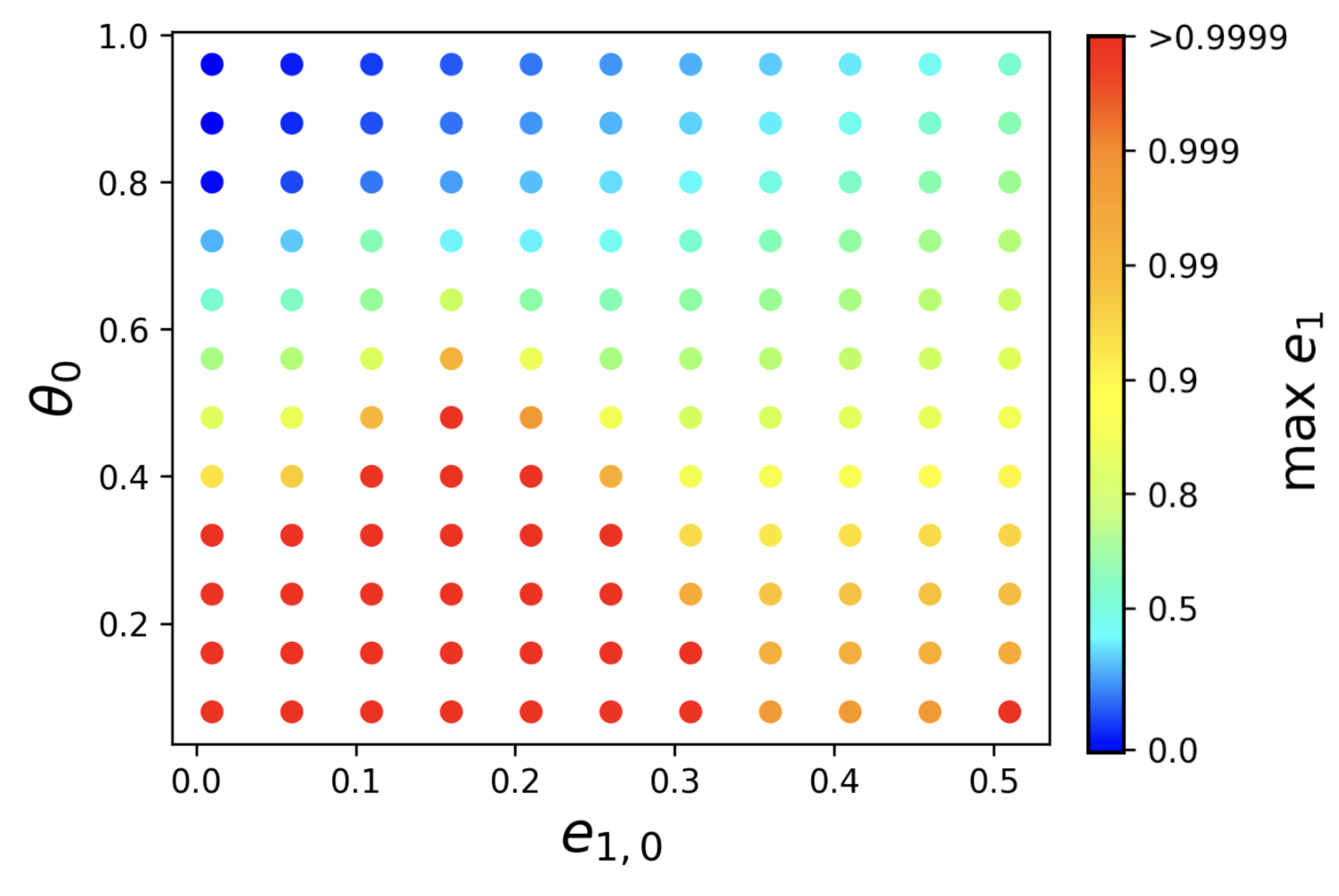}
\caption{\footnotesize Grid of initial eccentricities and inclinations ($\theta = \cos i$), where the color code corresponds to the maximum eccentricity obtained by integrating for 100$t_{\rm quad}$. We note that the color code is spaced non-linearly to observe features at both low and high eccentricities. Each system is set with $a_1 = 5$ au, $a_2 = 165$ au, and $e_2 = 0.5$, such that $\epsilon = 0.02$. Here, we set $m_1 = 1 M_{\odot}$, $m_2 = 1 M_J$, $m_3 = 1 M_{\odot}$, $\omega_{1,0} = 0^{\circ}$, and $\omega_{2,0} = 0^{\circ}$.} 
\label{fig:maxe_grid}
\end{center}
\end{figure}

For a circular orbit, the initial mutual inclination $i_0$ and the hierarchy parameter $\epsilon$ are the most important parameters for determining if EKL will lead a system to reach extreme eccentricities. Increasing the initial eccentricity $e_{1,0}$ of an orbit has two potential effects on the system. First, if the system does not have sufficient $i_0$ and $\epsilon$ to reach extreme eccentricities, the system may still "gently" oscillate, but the bounds of the oscillations will be higher than for initially circular orbits \citep[e.g.,][]{Lithwick+11,Teyssander+13,Katz+11}. Second, increasing the initial eccentricity may trigger the system to reach dramatically high eccentricities at lower inclinations than is possible on a circular orbit at fixed $\epsilon$ \citep{Li+14a,Li+14b}. This effect is strongest near $e_{1,0} \sim 0.15$, which is also roughly the value where the mean initial eccentricity in the simulations produces the best match to the observations.

These two effects are illustrated in Fig. \ref{fig:maxe_grid}. 
We construct a grid with $e_{1,0}$ varied from 0.01 to 0.51 and $\theta_0 = \cos i_0$ varied from 0.08 to 0.96. Each system is set with $a_1 = 5$ au, $a_2 = 165$ au, and $e_2 = 0.5$, such that $\epsilon = 0.02$. Here, we set $m_1 = 1 M_{\odot}$, $m_2 = 1 M_J$, $m_3 = 1 M_{\odot}$, $\omega_{1,0} = 0^{\circ}$, and $\omega_{2,0} = 0^{\circ}$. Each system is integrated for $100t_{\rm quad}$ including octupole-level secular effects and general relativity, but no tides. 

Fig. \ref{fig:maxe_grid} shows the grid of points with different initial eccentricities and inclinations, where the color code shows the maximum eccentricity obtained. At low inclinations ($\theta_0 \gtrsim 0.6$), the systems oscillate up to eccentricities of $\sim0.8$. However, increasing the initial eccentricity not only starts the system with an initially higher eccentricity, but leads the system to reach higher eccentricities than are possible on circular orbits. For example, at $\theta_0 \sim 1$, with $e_{1,0} = 0.01$, the system does not undergo significant eccentricity oscillations. At the same value of $\theta_0$, with $e_{1,0} \sim 0.5$ instead, the system can reach a maximum eccentricity of up to $\sim0.7$.

At higher inclinations ($\theta_0 \lesssim 0.6$), Fig. \ref{fig:maxe_grid} reveals an additional structure. This structure shows that for some fixed $\theta_0$, a slightly higher initial eccentricity can lead the system to dramatically more extreme eccentricities. For example, at $\theta_0 \sim 0.5$, the initially circular orbit has a maximum eccentricity near 0.8. However, if the same system begins with $e_{1,0} \approx 0.1$, the system achieves a maximum eccentricity greater than 0.99. These more extreme eccentricities not only produce a different shape in the eccentricity distribution, the smaller associated pericenter distances may lead to tidal disruptions or tidal migration. This structure shifts to even lower inclinations for increasing values of $\epsilon$, as noted before in e.g., \cite{Katz+11}, \cite{Lithwick+11}, \cite{Teyssandier+13}, and \cite{Li+14b}, . However, its application to the cold Jupiter eccentricities was missed.

The combination of the two aforementioned effects is responsible for the large differences in the resultant eccentricity distribution of the cold Jupiters between the case of initially circular orbits and with slightly eccentric orbits. At slightly higher initial eccentricities, EKL causes systems to oscillate at higher eccentricities, and in some cases, may lower the inclination needed for EKL to drive the system to extremely high eccentricities. Only a small amount of initial eccentricity is needed to achieve these effects, and as discussed below, we expect scattering to generate the necessary small amounts of initial eccentricity. 

\subsection{Comparison with other works}

Next, we compare the eccentricity distribution produced by EKL in this work with other studies. These studies employed slightly different cuts on their planet mass and semi-major axis ranges, but the overall goal of each was to explain the giant planet eccentricity distribution, excluding the close-in hot Jupiters. 

\begin{figure}
\begin{center}

\includegraphics[width=3.3in]{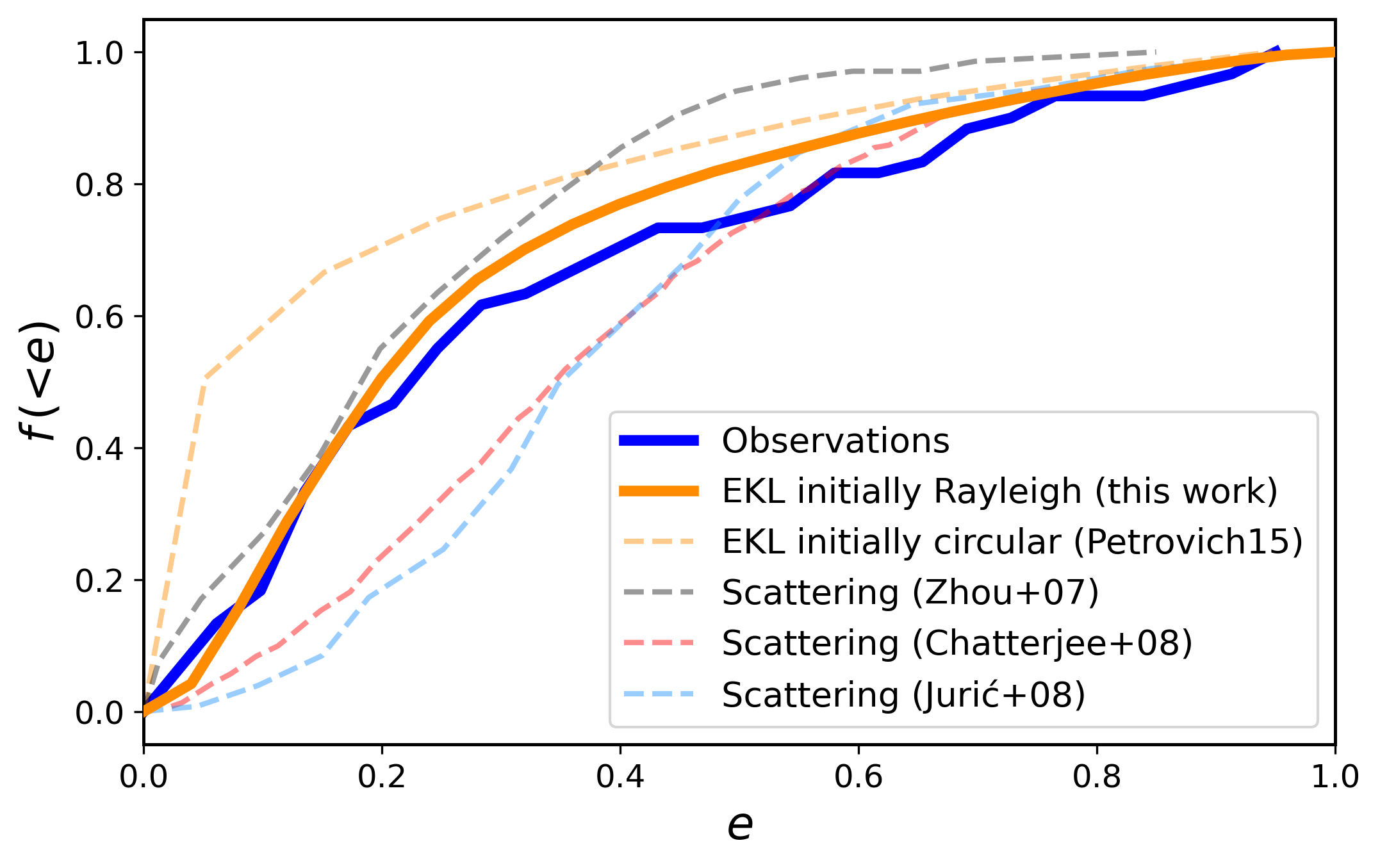}
\caption{\footnotesize Comparison of the eccentricity CDF between the observed MSSP sample (solid blue line) and the \texttt{r0.13} simulation run in this work (solid orange line). We also show, from low to high median eccentricity, the distribution produced by EKL with circular orbits in \cite{Petrovich15b} (dashed orange line), the scattering distribution in \cite{Zhou+07} (dashed grey line), the scattering distribution in \cite{Chatterjee+08} (solid red line), and the scattering distribution in \cite{Juric+08} (dashed light blue line).} 
\label{fig:cdf_comparison}
\end{center}
\end{figure}

In Fig. \ref{fig:cdf_comparison}, we show the cumulative distribution function (CDF) for the observed MSSP sample (solid blue line) and our \texttt{r0.13} simulation run eccentricity distribution (solid orange line). We compare to the EKL distribution with initially circular orbits from \cite{Petrovich15b} (dashed orange line), as well as scattering distributions from \cite{Zhou+07} (dashed grey line), \cite{Chatterjee+08} (dashed red line), and \cite{Juric+08} (dashed light blue line, their model \texttt{c50s05}). Each of these authors note that the mean eccentricity produced depends sensitively on the observationally ill-constrained primordial configuration of the planetary systems, but that the expected outcome of the scattering process is a Rayleigh distribution for a wide range of initial conditions. 

The present-day observed eccentricity distribution has a peak near $e = 0.2$ and a tail extending to $e\sim1$, which is not well-described by a Rayleigh distribution. A Rayleigh distribution tends to have either too few highly eccentric orbits, as seen in the CDF for \cite{Zhou+07}, or too few low eccentricity orbits, as seen in the CDFs for \cite{Juric+08} and \cite{Chatterjee+08}, and as described by these authors. Performing a Rayleigh distribution fit to the observed sample yields a KS-test $p$-value of $\sim0.01$ for the best fit, indicating that it is indeed inconsistent with the data. 

From Fig. \ref{fig:cdf_comparison}, we see that the EKL mechanism in this work with an \textit{initially} Rayleigh distribution of eccentricities provides a close match to the observations, as it able to reproduce both the peak at low eccentricities and the tail of the distribution out to high eccentricities. Our simulations have shown that the initial Rayleigh distribution must have a modest initial mean eccentricity ($\sim0.1-0.2$). In simulations of scattering \cite[e.g.,][]{Zhou+07,Juric+08,Ford+08,Raymond+10,Carrera+19},
lower eccentricity distributions tend to be produced when planets have low multiplicities or are more distantly spaced, when planet masses are lower or more even, or when there is eccentricity damping from a planetesimal disk. Alternatively, there may be a population of dynamically inactive planets with low eccentricity, and an active population with high eccentricity following the epoch of scattering. Furthermore, resonances in protoplanetary disks may themselves be a source of small initial eccentricities \citep[e.g.,][]{Ragusa+18,Li+23}. Future observational constraints on the primordial configurations of planetary systems may test whether the expected scattering outcome is consistent with the initial conditions in this work.

\subsection{Predictions for inclinations and obliquities}

\begin{figure*}[ht]
\begin{center}

\includegraphics[width=7.0in]{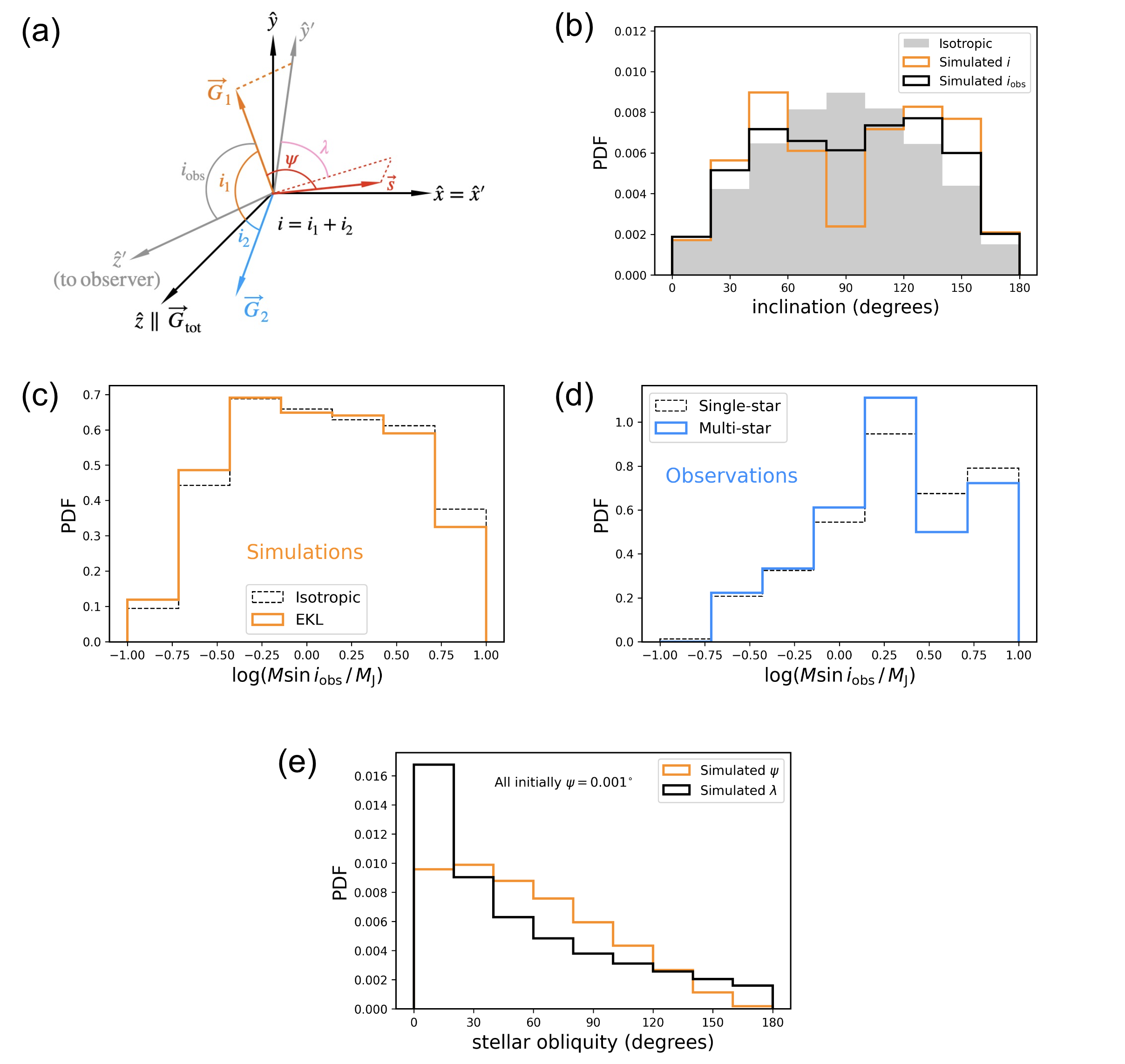}
\caption{\footnotesize Panel (a): Diagram (not to scale) illustrating important angles. In this example, the coordinate system $\hat{x}, \hat{y}, \hat{z}$ is defined by the total angular momentum of the system ($\vec{G}_{\rm tot} \parallel \hat{z}$) and is related to the observer frame $\hat{x}',\hat{y}',\hat{z}'$ by a rotation about the $\hat{x} = \hat{x}'$ axis. The angular momentum of the inner (outer) orbit is shown as the vector $\vec{G_1}$ ($\vec{G_2}$), and the respective angle with the $\hat{z}$-axis is $i_1$ ($i_2$). The mutual inclination $i$ is the angle between $\vec{G_1}$ and $\vec{G_2}$. The inclination of $G_1$ with respect to the line-of-sight is $i_{\rm obs}$. The stellar obliquity $\psi$ is the angle between the spin vector $\vec{s}$ of the star and $\vec{G_1}$. The projection of the obliquity onto the sky plane is $\lambda$. Panel (b): Isotropic distribution of inclinations (solid grey region), distribution of mutual inclinations in the \texttt{r0.13} simulation run (solid orange line), and distribution of $i_{\rm obs}$ (solid black line) calculated by rotating the simulations into the observer frame. Panel (c): Planetary mass in the \texttt{r0.13} simulations. We show the line-of-sight masses assuming an isotropic inclination distribution (dashed black line), and the line-of-sight masses produced by EKL (solid orange line). Panel (d): Observed line-of-sight masses for planets in the SSSP (dashed black line) and MSSP (solid blue line) samples. Panel (e): Distribution of stellar obliquities $\psi$ produced by EKL in the \texttt{r0.13} simulation run (solid orange line) and the sky-projected obliquities $\lambda$ (solid black line). All systems have spin-orbit angles initially aligned.} 
\label{fig:inc_distribution}
\end{center}
\end{figure*}

We now examine the distribution of mutual inclinations and stellar obliquities in the cold Jupiter population produced by stellar EKL. While the orbit of the stellar binary is randomly oriented with respect to an observer on Earth, EKL leads the planetary orbit to prefer certain inclinations within the triple system. Therefore, we wish to assess the impact of this mechanism on the expected distribution of observed line-of-sight masses, which usually presumes that
the inclinations are distributed isotropically with respect to the observer.

In panel (a) of Fig. \ref{fig:inc_distribution}, we show an illustrative diagram of important angles (not to scale). The coordinate system $\hat{x}, \hat{y}, \hat{z}$ is defined by the total angular momentum of the system ($\vec{G}_{\rm tot} \parallel \hat{z}$). In this example, the observer coordinate system $\hat{x}',\hat{y}',\hat{z}'$ is related to the total angular momentum frame by a rotation about the $\hat{x} = \hat{x}'$ axis. The angular momentum of the inner (outer) orbit is shown as the vector $\vec{G_1}$ ($\vec{G_2}$), and the respective angle with the $\hat{z}$-axis is $i_1$ ($i_2$). The mutual inclination $i = i_1 + i_2$ is the angle between $\vec{G_1}$ and $\vec{G_2}$. We note that $\vec{G_1}$, $\vec{G_2}$, and $\vec{G_{\rm tot}}$ lie in the same plane. For radial velocity measurements, mass can be measured up to a factor of $\sin i_{\rm obs}$, where $i_{\rm obs}$ is the angle between the angular momentum of the planet's orbit and the observer line-of-sight. The stellar obliquity $\psi$ is the angle between the spin vector $\vec{s}$ of the star and the inner orbit's angular momentum. The projection of the obliquity onto the sky plane is $\lambda$.

In panel (b) of Fig. \ref{fig:inc_distribution}, we show the mutual inclination distribution of cold Jupiters in the \texttt{r0.13} simulation run (solid orange line). An isotropic distribution is shown as the solid grey region for comparison. In the simulations, there is a bimodal distribution of mutual inclinations with peaks near $55^{\circ}$ and $125^{\circ}$. The paucity of inclinations near $90^{\circ}$ is attributed to the fact that at these high mutual inclinations, EKL tends to produce more extreme eccentricity oscillations, leading planets to be tidally disrupted or become hot Jupiters, both of which spend less time in the cold Jupiter distribution. Most systems that are observed as cold Jupiters are undergoing "gentle" oscillations near one of the modes of the distribution. To determine the distribution of $i_{\rm obs}$ that an observer would measure, we take 10000 random samples of $i_1$ from the simulations and calculate the distribution of $i_{\rm obs}$ by randomly rotating from the frame defined by the total angular momentum to the observer frame \citep[considering rotations in inclination, argument of periapse $\omega$, and longitude of ascending node $\Omega$, see for procedure][]{Murray+99}. See also e.g., \cite{Czekala+19} and \cite{Dupuy+22} for geometry in planet-hosting binaries. The resulting distribution of $i_{\rm obs}$ is shown as the solid black line in panel (b) of Fig. \ref{fig:inc_distribution}. The randomization of the binary orbit with respect to the observer dilutes but does not altogether erase the bimodality.

Next, we explore how this anisotropic distribution of inclinations affects radial velocity mass measurements. In panel (c) of Fig. \ref{fig:inc_distribution}, we compute the distribution of $m_2 \sin i_{\rm obs}$, considering both an isotropic distribution of inclinations with respect to the line-of-sight (dashed black line) and the distribution produced by EKL (solid orange line). We show the logarithm of the projected mass in Jupiter masses for planets with $0.1 M_J < m_2 \sin i_{\rm obs} < 10 M_J$. On average, the relative lack of orbits near $90^{\circ}$ leads masses to be lower by $\sim7$\% at the population level.

If EKL indeed produces deviations from isotropy, the observed distribution of line-of-sight masses should differ between the single-star and multi-star samples. These distributions are respectively shown as the dashed black line and solid blue line in panel (d) of Fig. \ref{fig:inc_distribution}. The lack of lower-mass planets relative to the simulations is due to observation bias for larger planets; here, we are interested in the shift between the single-star and multi-star distributions, which should be equally affected by this bias. The median of the multi-star mass distribution is $\sim25$\% lower than the single-star case, which is a discrepancy in the predicted direction. However, a KS-test suggests that the differences between the distributions are not statistically significant ($p \sim 0.3$). Similar to the lack of statistical significance between the single and multi-star eccentricity distributions, the single-star sample may be somewhat diluted by undetected companions. More observations of cold Jupiters will improve statistical sampling, and future observations that confirm or reject the presence of companions will help test these predictions more robustly. We note that we exclude the multi-planet systems from this analysis because of the complicated interplay between star-planet and planet-planet interactions that may affect the inclination distribution \citep[e.g.,][]{Denham+19,Wei+21}. Furthermore, there may be detection biases between inner and outer planets, and planet mass may be correlated to formation distance \citep[e.g.,][]{Mordasini+09}, which may swamp the small effect from stellar EKL.

Lastly, in panel (e) of Fig. \ref{fig:inc_distribution}, we show the distribution of the stellar obliquity $\psi$ (orange line) and sky-projected obliquity $\lambda$ (black line) for cold Jupiters\footnote{We note that our $\psi$ distribution falls off more steeply than others in the literature, such as \cite{Petrovich15b} and \cite{Stephan+18}. These studies fixed the planet's mass to be $1$~M$_J$, whereas the planet's mass is varied in this study. See \cite{Storch+15} for more on how planetary mass affects spin-orbit misalignment.}. We start all of the simulations with the spin and orbit initially aligned. The sky projections are calculated following the formalism in \cite{Fabrycky+09}. The distribution of $\lambda$ peaks near low values, but obliquities up to $\sim180^{\circ}$ are possible. A growing sample of warm Jupiter obliquities has been measured with the Rossiter-McLaughlin effect, with a preference toward alignment in single-star systems and some multi-star systems exhibiting significant misalignments \citep[e.g.,][]{Rice+22}. HD 80606 b, at 0.46 au, has $\lambda$ $\sim50^{\circ}$ \citep{Pont+09}. The recently discovered giant TIC 241249530 b at 0.64 au has $\lambda$ $\sim163^{\circ}$ \citep{Gupta+24}. Both of these obliquities are capable of being produced via EKL. While these systems do not fall under our "cold" definition, they are near the boundary and may be considered cold by other authors. Future measurements of more distant cold Jupiter obliquities will allow for comparisons with our results.

\section{Conclusion}
\label{sec:conclusion}

This paper focuses on the cold Jupiter population, heuristically defined as Jupiters with separations larger than 0.8 au, as shown in Fig. \ref{fig:observed_jupiters}. Specifically, we study the eccentricity distribution of cold Jupiters shaped by the EKL mechanism in stellar binaries. We find that the stellar EKL mechanism is able to reproduce the observations with statistical significance, as shown in Fig. \ref{fig:ecc_distributions}. KS-tests comparing the simulated and observed distributions give $p$-values up to $\sim0.5$, suggesting that these data are not inconsistent with originating from the same underlying distribution. 

Previous studies, such as those by \cite{Takeda+05} and \cite{Petrovich15b} found that if planets begin on initially circular orbits, EKL overproduces cold Jupiters at low eccentricities and underproduces cold Jupiters at moderate to high eccentricities. As shown in Fig. \ref{fig:maxe_grid}, only a small amount of initial eccentricity is needed to trigger the EKL mechanism to lead a system to higher eccentricities \citep[see also][]{Lithwick+11,Katz+11,Li+14b}. In this work, we examined a range of initial eccentricity distributions and found that a Rayleigh distribution with an average of $\sim0.1-0.2$ yields an agreement with the observed distribution, as shown in Fig. \ref{fig:ecc_distributions}. Small eccentricities are naturally produced in the early stages of planet-planet scattering \citep[e.g.,][]{Zhou+07,Juric+08,Chatterjee+08}. However, at the end of the dynamical evolution of planet-planet scattering, the final eccentricity distribution is inconsistent with observations, as Fig. \ref{fig:cdf_comparison} highlights.

The observed eccentricity distributions with different planetary and stellar multiplicities also provide evidence for the action of stellar EKL. The single-star observed sample is consistent with, but shifted to lower eccentricity values when compared to the multi-star sample, as highlighted by panel (b) in Fig. \ref{fig:ecc_distributions}. This result suggests that some fraction of undetected companion stars may be present in the single-star sample. By mixing simulations of non-EKL and EKL components, we estimate that roughly $40$\%-$50$\% of systems in the sample have stellar companions, although we note some high eccentricities may also be driven by undetected planetary companions or unbound stellar perturbers. Future observations to confirm or reject the presence of stellar companions will help test these results. The multi-planet control sample is shifted to much lower eccentricities and is statistically inconsistent with the single-planet sample according to the AD-test. While lower eccentricities in multi-planet systems have been noted before and attributed to dynamical stability \citep[e.g.,][]{Wright+09}, this result suggests that the presence of a planetary companion also hinders the EKL oscillations from any stellar perturbers. 

We also calculate predictions for the inclination and stellar obliquity distributions in the cold Jupiter population produced by EKL, as shown in Fig. \ref{fig:inc_distribution}. While the stellar binary orbit is random with respect to an observer, the planetary orbit prefers certain values. A relative dearth of mutual inclinations near $90^{\circ}$ leads the distribution of radial velocity mass estimates to be lower by $\sim7$\% compared to an isotropic distribution. There are hints of a difference between the measured masses of planets in single-star and multi-star systems, but more observations are needed to improve sampling and determine true stellar multiplicities. Future measurements of cold Jupiter obliquities will also test the results in this work.

Our work demonstrates that EKL plays a more important role in shaping the cold Jupiter eccentricity distribution than original studies suggest and, as a consequence, may play a more significant role in hot Jupiter formation. Within this framework, giant planets are born out of protoplanetary disks on initially near-circular orbits and undergo a period of planet-planet scattering on $\sim$Myr timescales. This period of scattering generates a moderate amount of initial eccentricity in the planetary orbits. Over much longer timescales ($\sim$Myr-Gyr), the EKL mechanism induces eccentricity oscillations and produces the observed eccentricity distribution that we see today. 
\\
\\
The authors gratefully acknowledge the anonymous referee for useful comments. The authors thank Erik Petigura, Yasuhiro Hasegawa, Yubo Su, Songhu Wang, Claire Williams, and Isabel Angelo for useful discussions. We acknowledge the support of NASA XRP grant 80NSSC23K0262. S. N. thanks Howard and Astrid Preston for their generous support. The authors also acknowledge the use of the UCLA cluster \textit{Hoffman2} for computational resources. This research has made use of NASA’s Astrophysics Data System Bibliographic Services. This research has also made use of the NASA Exoplanet Archive, which is operated by the California Institute of Technology, under contract with the National Aeronautics and Space Administration under the Exoplanet Exploration Program.

\software{
    NumPy \citep{numpy},
    SciPy \citep{scipy},
    Matplotlib \citep{matplotlib},
    Mathematica \citep{Mathematica},
    WebPlotDigitizer \citep{WebPlotDigitizer}
}

\newpage

\bibliography{paperbib, softwarebib}{}

\begin{thebibliography}{}
\expandafter\ifx\csname natexlab\endcsname\relax\def\natexlab#1{#1}\fi
\providecommand{\url}[1]{\href{#1}{#1}}
\providecommand{\dodoi}[1]{doi:~\href{http://doi.org/#1}{\nolinkurl{#1}}}
\providecommand{\doeprint}[1]{\href{http://ascl.net/#1}{\nolinkurl{http://ascl.net/#1}}}
\providecommand{\doarXiv}[1]{\href{https://arxiv.org/abs/#1}{\nolinkurl{https://arxiv.org/abs/#1}}}

\bibitem[{{Adams} \& {Laughlin}(2003)}]{Adams+03}
{Adams}, F.~C., \& {Laughlin}, G. 2003, \icarus, 163, 290, \dodoi{10.1016/S0019-1035(03)00081-2}

\bibitem[{{Akeson} {et~al.}(2013){Akeson}, {Chen}, {Ciardi}, {Crane}, {Good}, {Harbut}, {Jackson}, {Kane}, {Laity}, {Leifer}, {Lynn}, {McElroy}, {Papin}, {Plavchan}, {Ram{\'\i}rez}, {Rey}, {von Braun}, {Wittman}, {Abajian}, {Ali}, {Beichman}, {Beekley}, {Berriman}, {Berukoff}, {Bryden}, {Chan}, {Groom}, {Lau}, {Payne}, {Regelson}, {Saucedo}, {Schmitz}, {Stauffer}, {Wyatt}, \& {Zhang}}]{Akeson+13}
{Akeson}, R.~L., {Chen}, X., {Ciardi}, D., {et~al.} 2013, \pasp, 125, 989, \dodoi{10.1086/672273}

\bibitem[{{Anderson} {et~al.}(2016){Anderson}, {Storch}, \& {Lai}}]{Anderson+16}
{Anderson}, K.~R., {Storch}, N.~I., \& {Lai}, D. 2016, \mnras, 456, 3671, \dodoi{10.1093/mnras/stv2906}

\bibitem[{{Angelo} {et~al.}(2022){Angelo}, {Naoz}, {Petigura}, {MacDougall}, {Stephan}, {Isaacson}, \& {Howard}}]{Angelo+22}
{Angelo}, I., {Naoz}, S., {Petigura}, E., {et~al.} 2022, \aj, 163, 227, \dodoi{10.3847/1538-3881/ac6094}

\bibitem[{{Antognini}(2015)}]{Antognini15}
{Antognini}, J.~M.~O. 2015, \mnras, 452, 3610, \dodoi{10.1093/mnras/stv1552}

\bibitem[{{Carrera} {et~al.}(2019){Carrera}, {Raymond}, \& {Davies}}]{Carrera+19}
{Carrera}, D., {Raymond}, S.~N., \& {Davies}, M.~B. 2019, \aap, 629, L7, \dodoi{10.1051/0004-6361/201935744}

\bibitem[{{Chatterjee} {et~al.}(2008){Chatterjee}, {Ford}, {Matsumura}, \& {Rasio}}]{Chatterjee+08}
{Chatterjee}, S., {Ford}, E.~B., {Matsumura}, S., \& {Rasio}, F.~A. 2008, \apj, 686, 580, \dodoi{10.1086/590227}

\bibitem[{{Cumming}(2004)}]{Cumming+04}
{Cumming}, A. 2004, \mnras, 354, 1165, \dodoi{10.1111/j.1365-2966.2004.08275.x}

\bibitem[{{Czekala} {et~al.}(2019){Czekala}, {Chiang}, {Andrews}, {Jensen}, {Torres}, {Wilner}, {Stassun}, \& {Macintosh}}]{Czekala+19}
{Czekala}, I., {Chiang}, E., {Andrews}, S.~M., {et~al.} 2019, \apj, 883, 22, \dodoi{10.3847/1538-4357/ab287b}

\bibitem[{{Dawson} \& {Johnson}(2018)}]{Dawson+18}
{Dawson}, R.~I., \& {Johnson}, J.~A. 2018, \araa, 56, 175, \dodoi{10.1146/annurev-astro-081817-051853}

\bibitem[{{Denham} {et~al.}(2019){Denham}, {Naoz}, {Hoang}, {Stephan}, \& {Farr}}]{Denham+19}
{Denham}, P., {Naoz}, S., {Hoang}, B.-M., {Stephan}, A.~P., \& {Farr}, W.~M. 2019, \mnras, 482, 4146, \dodoi{10.1093/mnras/sty2830}

\bibitem[{{Dupuy} {et~al.}(2022){Dupuy}, {Kraus}, {Kratter}, {Rizzuto}, {Mann}, {Huber}, \& {Ireland}}]{Dupuy+22}
{Dupuy}, T.~J., {Kraus}, A.~L., {Kratter}, K.~M., {et~al.} 2022, \mnras, 512, 648, \dodoi{10.1093/mnras/stac306}

\bibitem[{{Duquennoy} \& {Mayor}(1991)}]{Duquennoy+91}
{Duquennoy}, A., \& {Mayor}, M. 1991, \aap, 248, 485

\bibitem[{{Eggleton} {et~al.}(1998){Eggleton}, {Kiseleva}, \& {Hut}}]{Eggleton98}
{Eggleton}, P.~P., {Kiseleva}, L.~G., \& {Hut}, P. 1998, \apj, 499, 853, \dodoi{10.1086/305670}

\bibitem[{{Eggleton} \& {Kiseleva-Eggleton}(2001)}]{Eggleton+01}
{Eggleton}, P.~P., \& {Kiseleva-Eggleton}, L. 2001, \apj, 562, 1012, \dodoi{10.1086/323843}

\bibitem[{{Fabrycky} \& {Tremaine}(2007)}]{Fabrycky+07}
{Fabrycky}, D., \& {Tremaine}, S. 2007, \apj, 669, 1298, \dodoi{10.1086/521702}

\bibitem[{{Fabrycky} \& {Winn}(2009)}]{Fabrycky+09}
{Fabrycky}, D.~C., \& {Winn}, J.~N. 2009, \apj, 696, 1230, \dodoi{10.1088/0004-637X/696/2/1230}

\bibitem[{{Faridani} {et~al.}(2022){Faridani}, {Naoz}, {Wei}, \& {Farr}}]{Faridani+22}
{Faridani}, T.~H., {Naoz}, S., {Wei}, L., \& {Farr}, W.~M. 2022, \apj, 932, 78, \dodoi{10.3847/1538-4357/ac6e38}

\bibitem[{{Fischer} \& {Marcy}(1992)}]{Fischer+92}
{Fischer}, D.~A., \& {Marcy}, G.~W. 1992, \apj, 396, 178, \dodoi{10.1086/171708}

\bibitem[{{Ford} {et~al.}(2000){Ford}, {Kozinsky}, \& {Rasio}}]{Ford+00}
{Ford}, E.~B., {Kozinsky}, B., \& {Rasio}, F.~A. 2000, \apj, 535, 385, \dodoi{10.1086/308815}

\bibitem[{{Ford} {et~al.}(2005){Ford}, {Lystad}, \& {Rasio}}]{Ford+05}
{Ford}, E.~B., {Lystad}, V., \& {Rasio}, F.~A. 2005, \nat, 434, 873, \dodoi{10.1038/nature03427}

\bibitem[{{Ford} \& {Rasio}(2008)}]{Ford+08}
{Ford}, E.~B., \& {Rasio}, F.~A. 2008, \apj, 686, 621, \dodoi{10.1086/590926}

\bibitem[{{Gaudi} \& {Winn}(2007)}]{Gaudi+07}
{Gaudi}, B.~S., \& {Winn}, J.~N. 2007, \apj, 655, 550, \dodoi{10.1086/509910}

\bibitem[{{Guillochon} {et~al.}(2011){Guillochon}, {Ramirez-Ruiz}, \& {Lin}}]{Guillochon+11}
{Guillochon}, J., {Ramirez-Ruiz}, E., \& {Lin}, D. 2011, \apj, 732, 74, \dodoi{10.1088/0004-637X/732/2/74}

\bibitem[{{Gupta} {et~al.}(2024){Gupta}, {Millholland}, {Im}, {Dong}, {Jackson}, {Carleo}, {Libby-Roberts}, {Delamer}, {Giovinazzi}, {Lin}, {Kanodia}, {Wang}, {Stassun}, {Masseron}, {Dragomir}, {Mahadevan}, {Wright}, {Alvarado-Montes}, {Bender}, {Blake}, {Caldwell}, {Ca{\~n}as}, {Cochran}, {Dalba}, {Everett}, {Fernandez}, {Golub}, {Guillet}, {Halverson}, {Hebb}, {Higuera}, {Huang}, {Klusmeyer}, {Knight}, {Leroux}, {Logsdon}, {Loose}, {McElwain}, {Monson}, {Ninan}, {Nowak}, {Palle}, {Patel}, {Pepper}, {Primm}, {Rajagopal}, {Robertson}, {Roy}, {Schneider}, {Schwab}, {Schweiker}, {Sgro}, {Shimizu}, {Simard}, {Stef{\'a}nsson}, {Stevens}, {Villanueva}, {Wisniewski}, {Will}, \& {Ziegler}}]{Gupta+24}
{Gupta}, A.~F., {Millholland}, S.~C., {Im}, H., {et~al.} 2024, \nat, 632, 50, \dodoi{10.1038/s41586-024-07688-3}

\bibitem[{{Harrington}(1968)}]{Harrington68}
{Harrington}, R.~S. 1968, \aj, 73, 190, \dodoi{10.1086/110614}

\bibitem[{{Holman} {et~al.}(1997){Holman}, {Touma}, \& {Tremaine}}]{Holman+97}
{Holman}, M., {Touma}, J., \& {Tremaine}, S. 1997, \nat, 386, 254, \dodoi{10.1038/386254a0}

\bibitem[{Hunter(2007)}]{matplotlib}
Hunter, J.~D. 2007, Computing in Science \& Engineering, 9, 90, \dodoi{10.1109/MCSE.2007.55}

\bibitem[{{Hurley} {et~al.}(2000){Hurley}, {Pols}, \& {Tout}}]{Hurley2000}
{Hurley}, J.~R., {Pols}, O.~R., \& {Tout}, C.~A. 2000, \mnras, 315, 543, \dodoi{10.1046/j.1365-8711.2000.03426.x}

\bibitem[{{Jeans}(1919)}]{Jeans19}
{Jeans}, J.~H. 1919, \mnras, 79, 408, \dodoi{10.1093/mnras/79.6.408}

\bibitem[{{Jones} {et~al.}(2003){Jones}, {Butler}, {Tinney}, {Marcy}, {Penny}, {McCarthy}, \& {Carter}}]{Jones+03}
{Jones}, H. R.~A., {Butler}, R.~P., {Tinney}, C.~G., {et~al.} 2003, \mnras, 341, 948, \dodoi{10.1046/j.1365-8711.2003.06481.x}

\bibitem[{{Juri{\'c}} \& {Tremaine}(2008)}]{Juric+08}
{Juri{\'c}}, M., \& {Tremaine}, S. 2008, \apj, 686, 603, \dodoi{10.1086/590047}

\bibitem[{{Katz} {et~al.}(2011){Katz}, {Dong}, \& {Malhotra}}]{Katz+11}
{Katz}, B., {Dong}, S., \& {Malhotra}, R. 2011, \prl, 107, 181101, \dodoi{10.1103/PhysRevLett.107.181101}

\bibitem[{{Kipping}(2013)}]{Kipping13}
{Kipping}, D.~M. 2013, \mnras, 434, L51, \dodoi{10.1093/mnrasl/slt075}

\bibitem[{{Klein} \& {Katz}(2024)}]{Klein+24}
{Klein}, Y.~Y., \& {Katz}, B. 2024, \aj, 167, 80, \dodoi{10.3847/1538-3881/ad18b6}

\bibitem[{{Knutson} {et~al.}(2014){Knutson}, {Fulton}, {Montet}, {Kao}, {Ngo}, {Howard}, {Crepp}, {Hinkley}, {Bakos}, {Batygin}, {Johnson}, {Morton}, \& {Muirhead}}]{Knutson+14}
{Knutson}, H.~A., {Fulton}, B.~J., {Montet}, B.~T., {et~al.} 2014, \apj, 785, 126, \dodoi{10.1088/0004-637X/785/2/126}

\bibitem[{{Kozai}(1962)}]{Kozai62}
{Kozai}, Y. 1962, \aj, 67, 591, \dodoi{10.1086/108790}

\bibitem[{{Li} {et~al.}(2014{\natexlab{a}}){Li}, {Naoz}, {Holman}, \& {Loeb}}]{Li+14a}
{Li}, G., {Naoz}, S., {Holman}, M., \& {Loeb}, A. 2014{\natexlab{a}}, \apj, 791, 86, \dodoi{10.1088/0004-637X/791/2/86}

\bibitem[{{Li} {et~al.}(2014{\natexlab{b}}){Li}, {Naoz}, {Kocsis}, \& {Loeb}}]{Li+14b}
{Li}, G., {Naoz}, S., {Kocsis}, B., \& {Loeb}, A. 2014{\natexlab{b}}, \apj, 785, 116, \dodoi{10.1088/0004-637X/785/2/116}

\bibitem[{{Li} \& {Lai}(2023)}]{Li+23}
{Li}, J., \& {Lai}, D. 2023, \apj, 956, 17, \dodoi{10.3847/1538-4357/aced89}

\bibitem[{{Lidov}(1962)}]{Lidov62}
{Lidov}, M.~L. 1962, \planss, 9, 719, \dodoi{10.1016/0032-0633(62)90129-0}

\bibitem[{{Lithwick} \& {Naoz}(2011)}]{Lithwick+11}
{Lithwick}, Y., \& {Naoz}, S. 2011, \apj, 742, 94, \dodoi{10.1088/0004-637X/742/2/94}

\bibitem[{{Malmberg} \& {Davies}(2009)}]{Malmberg+09}
{Malmberg}, D., \& {Davies}, M.~B. 2009, \mnras, 394, L26, \dodoi{10.1111/j.1745-3933.2008.00603.x}

\bibitem[{{Marcy} \& {Butler}(2000)}]{Marcy+2000}
{Marcy}, G.~W., \& {Butler}, R.~P. 2000, \pasp, 112, 137, \dodoi{10.1086/316516}

\bibitem[{{Mardling} \& {Aarseth}(2001)}]{Mardling+01}
{Mardling}, R.~A., \& {Aarseth}, S.~J. 2001, \mnras, 321, 398, \dodoi{10.1046/j.1365-8711.2001.03974.x}

\bibitem[{{Moe} \& {Di Stefano}(2017)}]{Moe+17}
{Moe}, M., \& {Di Stefano}, R. 2017, \apjs, 230, 15, \dodoi{10.3847/1538-4365/aa6fb6}

\bibitem[{{Moorhead} {et~al.}(2011){Moorhead}, {Ford}, {Morehead}, {Rowe}, {Borucki}, {Batalha}, {Bryson}, {Caldwell}, {Fabrycky}, {Gautier}, {Koch}, {Holman}, {Jenkins}, {Li}, {Lissauer}, {Lucas}, {Marcy}, {Quinn}, {Quintana}, {Ragozzine}, {Shporer}, {Still}, \& {Torres}}]{Moorhead+11}
{Moorhead}, A.~V., {Ford}, E.~B., {Morehead}, R.~C., {et~al.} 2011, \apjs, 197, 1, \dodoi{10.1088/0067-0049/197/1/1}

\bibitem[{{Mordasini} {et~al.}(2009){Mordasini}, {Alibert}, {Benz}, \& {Naef}}]{Mordasini+09}
{Mordasini}, C., {Alibert}, Y., {Benz}, W., \& {Naef}, D. 2009, \aap, 501, 1161, \dodoi{10.1051/0004-6361/200810697}

\bibitem[{{Murray} \& {Dermott}(1999)}]{Murray+99}
{Murray}, C.~D., \& {Dermott}, S.~F. 1999, {Solar System Dynamics}, \dodoi{10.1017/CBO9781139174817}

\bibitem[{{Nagasawa} \& {Ida}(2011)}]{Nagasawa+11}
{Nagasawa}, M., \& {Ida}, S. 2011, in AAS/Division for Extreme Solar Systems Abstracts, Vol.~2, AAS/Division for Extreme Solar Systems Abstracts, 6.06

\bibitem[{{Nagasawa} {et~al.}(2008){Nagasawa}, {Ida}, \& {Bessho}}]{Nagasawa+08}
{Nagasawa}, M., {Ida}, S., \& {Bessho}, T. 2008, \apj, 678, 498, \dodoi{10.1086/529369}

\bibitem[{{Naoz}(2016)}]{Naoz16}
{Naoz}, S. 2016, \araa, 54, 441, \dodoi{10.1146/annurev-astro-081915-023315}

\bibitem[{{Naoz} {et~al.}(2011){Naoz}, {Farr}, {Lithwick}, {Rasio}, \& {Teyssandier}}]{Naoz+11}
{Naoz}, S., {Farr}, W.~M., {Lithwick}, Y., {Rasio}, F.~A., \& {Teyssandier}, J. 2011, \nat, 473, 187, \dodoi{10.1038/nature10076}

\bibitem[{{Naoz} {et~al.}(2013{\natexlab{a}}){Naoz}, {Farr}, {Lithwick}, {Rasio}, \& {Teyssandier}}]{Naoz+13}
---. 2013{\natexlab{a}}, \mnras, 431, 2155, \dodoi{10.1093/mnras/stt302}

\bibitem[{{Naoz} {et~al.}(2012){Naoz}, {Farr}, \& {Rasio}}]{Naoz+12}
{Naoz}, S., {Farr}, W.~M., \& {Rasio}, F.~A. 2012, \apjl, 754, L36, \dodoi{10.1088/2041-8205/754/2/L36}

\bibitem[{{Naoz} {et~al.}(2016){Naoz}, {Fragos}, {Geller}, {Stephan}, \& {Rasio}}]{Naoz+16}
{Naoz}, S., {Fragos}, T., {Geller}, A., {Stephan}, A.~P., \& {Rasio}, F.~A. 2016, \apjl, 822, L24, \dodoi{10.3847/2041-8205/822/2/L24}

\bibitem[{{Naoz} {et~al.}(2013{\natexlab{b}}){Naoz}, {Kocsis}, {Loeb}, \& {Yunes}}]{Naoz+13b}
{Naoz}, S., {Kocsis}, B., {Loeb}, A., \& {Yunes}, N. 2013{\natexlab{b}}, \apj, 773, 187, \dodoi{10.1088/0004-637X/773/2/187}

\bibitem[{{NASA Exoplanet Archive}(2024)}]{PSCompPars}
{NASA Exoplanet Archive}. 2024, Planetary Systems Composite Parameters,  NExScI-Caltech/IPAC, \dodoi{10.26133/NEA13}

\bibitem[{{Ngo} {et~al.}(2015){Ngo}, {Knutson}, {Hinkley}, {Crepp}, {Bechter}, {Batygin}, {Howard}, {Johnson}, {Morton}, \& {Muirhead}}]{Ngo+15}
{Ngo}, H., {Knutson}, H.~A., {Hinkley}, S., {et~al.} 2015, \apj, 800, 138, \dodoi{10.1088/0004-637X/800/2/138}

\bibitem[{{Petrovich}(2015{\natexlab{a}})}]{Petrovich+15a}
{Petrovich}, C. 2015{\natexlab{a}}, \apj, 805, 75, \dodoi{10.1088/0004-637X/805/1/75}

\bibitem[{{Petrovich}(2015{\natexlab{b}})}]{Petrovich15b}
---. 2015{\natexlab{b}}, \apj, 799, 27, \dodoi{10.1088/0004-637X/799/1/27}

\bibitem[{{Pont} {et~al.}(2009){Pont}, {H{\'e}brard}, {Irwin}, {Bouchy}, {Moutou}, {Ehrenreich}, {Guillot}, {Aigrain}, {Bonfils}, {Berta}, {Boisse}, {Burke}, {Charbonneau}, {Delfosse}, {Desort}, {Eggenberger}, {Forveille}, {Lagrange}, {Lovis}, {Nutzman}, {Pepe}, {Perrier}, {Queloz}, {Santos}, {S{\'e}gransan}, {Udry}, \& {Vidal-Madjar}}]{Pont+09}
{Pont}, F., {H{\'e}brard}, G., {Irwin}, J.~M., {et~al.} 2009, \aap, 502, 695, \dodoi{10.1051/0004-6361/200912463}

\bibitem[{{Pu} \& {Lai}(2018)}]{Pu+18}
{Pu}, B., \& {Lai}, D. 2018, \mnras, 478, 197, \dodoi{10.1093/mnras/sty1098}

\bibitem[{{Raghavan} {et~al.}(2010){Raghavan}, {McAlister}, {Henry}, {Latham}, {Marcy}, {Mason}, {Gies}, {White}, \& {ten Brummelaar}}]{Raghavan+10}
{Raghavan}, D., {McAlister}, H.~A., {Henry}, T.~J., {et~al.} 2010, \apjs, 190, 1, \dodoi{10.1088/0067-0049/190/1/1}

\bibitem[{{Ragusa} {et~al.}(2018){Ragusa}, {Rosotti}, {Teyssandier}, {Booth}, {Clarke}, \& {Lodato}}]{Ragusa+18}
{Ragusa}, E., {Rosotti}, G., {Teyssandier}, J., {et~al.} 2018, \mnras, 474, 4460, \dodoi{10.1093/mnras/stx3094}

\bibitem[{{Rasio} \& {Ford}(1996)}]{Rasio+96}
{Rasio}, F.~A., \& {Ford}, E.~B. 1996, Science, 274, 954, \dodoi{10.1126/science.274.5289.954}

\bibitem[{{Raymond} {et~al.}(2010){Raymond}, {Armitage}, \& {Gorelick}}]{Raymond+10}
{Raymond}, S.~N., {Armitage}, P.~J., \& {Gorelick}, N. 2010, \apj, 711, 772, \dodoi{10.1088/0004-637X/711/2/772}

\bibitem[{{Rice} {et~al.}(2022){Rice}, {Wang}, {Wang}, {Stef{\'a}nsson}, {Isaacson}, {Howard}, {Logsdon}, {Schweiker}, {Dai}, {Brinkman}, {Giacalone}, \& {Holcomb}}]{Rice+22}
{Rice}, M., {Wang}, S., {Wang}, X.-Y., {et~al.} 2022, \aj, 164, 104, \dodoi{10.3847/1538-3881/ac8153}

\bibitem[{Rohatgi(2024)}]{WebPlotDigitizer}
Rohatgi, A. 2024, WebPlotDigitizer, 5.2.
\newblock \url{https://automeris.io}

\bibitem[{{Salpeter}(1955)}]{Salpeter55}
{Salpeter}, E.~E. 1955, \apj, 121, 161, \dodoi{10.1086/145971}

\bibitem[{{Shariat} {et~al.}(2024){Shariat}, {Naoz}, {El-Badry}, {Rodriguez}, {Hansen}, {Angelo}, \& {Stephan}}]{Shariat+24}
{Shariat}, C., {Naoz}, S., {El-Badry}, K., {et~al.} 2024, arXiv e-prints, arXiv:2407.06257, \dodoi{10.48550/arXiv.2407.06257}

\bibitem[{{Shariat} {et~al.}(2023){Shariat}, {Naoz}, {Hansen}, {Angelo}, {Michaely}, \& {Stephan}}]{Shariat+23}
{Shariat}, C., {Naoz}, S., {Hansen}, B. M.~S., {et~al.} 2023, \apjl, 955, L14, \dodoi{10.3847/2041-8213/acf76b}

\bibitem[{{Shen} \& {Turner}(2008)}]{Shen+08}
{Shen}, Y., \& {Turner}, E.~L. 2008, \apj, 685, 553, \dodoi{10.1086/590548}

\bibitem[{{Stephan} {et~al.}(2024){Stephan}, {Martin}, {Naoz}, {Hughes}, \& {Shariat}}]{Stephan+24}
{Stephan}, A.~P., {Martin}, D.~V., {Naoz}, S., {Hughes}, N.~R., \& {Shariat}, C. 2024, arXiv e-prints, arXiv:2408.13307, \dodoi{10.48550/arXiv.2408.13307}

\bibitem[{{Stephan} {et~al.}(2018){Stephan}, {Naoz}, \& {Gaudi}}]{Stephan+18}
{Stephan}, A.~P., {Naoz}, S., \& {Gaudi}, B.~S. 2018, \aj, 156, 128, \dodoi{10.3847/1538-3881/aad6e5}

\bibitem[{{Stephan} {et~al.}(2021){Stephan}, {Naoz}, \& {Gaudi}}]{Stephan21}
---. 2021, \apj, 922, 4, \dodoi{10.3847/1538-4357/ac22a9}

\bibitem[{{Stephan} {et~al.}(2020){Stephan}, {Naoz}, {Gaudi}, \& {Salas}}]{Stephan+20}
{Stephan}, A.~P., {Naoz}, S., {Gaudi}, B.~S., \& {Salas}, J.~M. 2020, \apj, 889, 45, \dodoi{10.3847/1538-4357/ab5b00}

\bibitem[{{Stephan} {et~al.}(2016){Stephan}, {Naoz}, {Ghez}, {Witzel}, {Sitarski}, {Do}, \& {Kocsis}}]{Stephan+16}
{Stephan}, A.~P., {Naoz}, S., {Ghez}, A.~M., {et~al.} 2016, \mnras, 460, 3494, \dodoi{10.1093/mnras/stw1220}

\bibitem[{{Stephan} {et~al.}(2019){Stephan}, {Naoz}, {Ghez}, {Morris}, {Ciurlo}, {Do}, {Breivik}, {Coughlin}, \& {Rodriguez}}]{Stephan+19}
---. 2019, \apj, 878, 58, \dodoi{10.3847/1538-4357/ab1e4d}

\bibitem[{{Storch} \& {Lai}(2015)}]{Storch+15}
{Storch}, N.~I., \& {Lai}, D. 2015, \mnras, 448, 1821, \dodoi{10.1093/mnras/stv119}

\bibitem[{{Takeda} \& {Rasio}(2005)}]{Takeda+05}
{Takeda}, G., \& {Rasio}, F.~A. 2005, \apj, 627, 1001, \dodoi{10.1086/430467}

\bibitem[{{Teyssandier} {et~al.}(2019){Teyssandier}, {Lai}, \& {Vick}}]{Teyssandier+19}
{Teyssandier}, J., {Lai}, D., \& {Vick}, M. 2019, \mnras, 486, 2265, \dodoi{10.1093/mnras/stz1011}

\bibitem[{{Teyssandier} {et~al.}(2013{\natexlab{a}}){Teyssandier}, {Naoz}, {Lizarraga}, \& {Rasio}}]{Teyssandier+13}
{Teyssandier}, J., {Naoz}, S., {Lizarraga}, I., \& {Rasio}, F.~A. 2013{\natexlab{a}}, \apj, 779, 166, \dodoi{10.1088/0004-637X/779/2/166}

\bibitem[{{Teyssandier} {et~al.}(2013{\natexlab{b}}){Teyssandier}, {Naoz}, {Lizarraga}, \& {Rasio}}]{Teyssander+13}
---. 2013{\natexlab{b}}, \apj, 779, 166, \dodoi{10.1088/0004-637X/779/2/166}

\bibitem[{{Udry} {et~al.}(2003){Udry}, {Mayor}, \& {Santos}}]{Udry+03}
{Udry}, S., {Mayor}, M., \& {Santos}, N.~C. 2003, \aap, 407, 369, \dodoi{10.1051/0004-6361:20030843}

\bibitem[{Van Der~Walt {et~al.}(2011)Van Der~Walt, Colbert, \& Varoquaux}]{numpy}
Van Der~Walt, S., Colbert, S.~C., \& Varoquaux, G. 2011, Computing in Science \& Engineering, 13, 22

\bibitem[{{Virtanen} {et~al.}(2020){Virtanen}, {Gommers}, {Oliphant}, {Haberland}, {Reddy}, {Cournapeau}, {Burovski}, {Peterson}, {Weckesser}, {Bright}, {van der Walt}, {Brett}, {Wilson}, {Jarrod Millman}, {Mayorov}, {Nelson}, {Jones}, {Kern}, {Larson}, {Carey}, {Polat}, {Feng}, {Moore}, {Vand erPlas}, {Laxalde}, {Perktold}, {Cimrman}, {Henriksen}, {Quintero}, {Harris}, {Archibald}, {Ribeiro}, {Pedregosa}, {van Mulbregt}, \& {Contributors}}]{scipy}
{Virtanen}, P., {Gommers}, R., {Oliphant}, T.~E., {et~al.} 2020, Nature Methods, 17, 261, \dodoi{https://doi.org/10.1038/s41592-019-0686-2}

\bibitem[{{Wei} {et~al.}(2021){Wei}, {Naoz}, {Faridani}, \& {Farr}}]{Wei+21}
{Wei}, L., {Naoz}, S., {Faridani}, T., \& {Farr}, W.~M. 2021, \apj, 923, 118, \dodoi{10.3847/1538-4357/ac2c70}

\bibitem[{{Weldon} {et~al.}(2024){Weldon}, {Naoz}, \& {Hansen}}]{Weldon+24}
{Weldon}, G.~C., {Naoz}, S., \& {Hansen}, B. M.~S. 2024, arXiv e-prints, arXiv:2405.20377, \dodoi{10.48550/arXiv.2405.20377}

\bibitem[{{Winn} \& {Fabrycky}(2015)}]{Winn+15}
{Winn}, J.~N., \& {Fabrycky}, D.~C. 2015, \araa, 53, 409, \dodoi{10.1146/annurev-astro-082214-122246}

\bibitem[{{Wolfram Research, Inc.}(2020)}]{Mathematica}
{Wolfram Research, Inc.} 2020, Mathematica 12.1.1.0.
\newblock \url{https://www.wolfram.com/mathematica}

\bibitem[{{Wright} {et~al.}(2009){Wright}, {Upadhyay}, {Marcy}, {Fischer}, {Ford}, \& {Johnson}}]{Wright+09}
{Wright}, J.~T., {Upadhyay}, S., {Marcy}, G.~W., {et~al.} 2009, \apj, 693, 1084, \dodoi{10.1088/0004-637X/693/2/1084}

\bibitem[{{Wu} \& {Lithwick}(2011)}]{Wu+11}
{Wu}, Y., \& {Lithwick}, Y. 2011, \apj, 735, 109, \dodoi{10.1088/0004-637X/735/2/109}

\bibitem[{{Wu} \& {Murray}(2003)}]{Wu+03}
{Wu}, Y., \& {Murray}, N. 2003, \apj, 589, 605, \dodoi{10.1086/374598}

\bibitem[{{Zhou} {et~al.}(2007){Zhou}, {Lin}, \& {Sun}}]{Zhou+07}
{Zhou}, J.-L., {Lin}, D. N.~C., \& {Sun}, Y.-S. 2007, \apj, 666, 423, \dodoi{10.1086/519918}

\end{thebibliography}
\bibliographystyle{aasjournal}

\appendix

\section{Warm/Cold Boundary}
\label{app:WC_boundary}

Various works have set the boundary between warm and cold Jupiters from $\sim0.4-1$ au. Here, we physically motivate our choice of 0.8 au. We consider cold Jupiters to be on average those planets that that are coupled to their stellar perturber, whereas warm Jupiters are planets that have decoupled and are in transit to becoming hot Jupiters. This decoupling occurs when the general relativistic precession timescale \citep[e.g.,][]{Naoz+13b, Petrovich15b} becomes lower than the timescale of EKL oscillations \citep[e.g.,][]{Antognini15}, which happens at a semi-major axis of 
\begin{equation}
    a_{1} \approx 0.45 \, {\rm au} \left( \frac{m_1}{M_{\odot}} \right)^{1/2} \left( \frac{m_3}{M_{\odot}} \right)^{-1/4} \left( \frac{a_2 \sqrt{1-e_2^2}}{100 \, {\rm au}} \right)^{3/4} \, .
\end{equation}
For the population of stellar perturbers in this work (see \S \ref{subsec:inicons}), the median value at which decoupling occurs is $\sim0.8$ au. This corresponds to Sun-like stars separated by $a_2 = 250$ au with $e_2 = 0.5$.

\section{Example time evolution}
\label{app:time_evolution}

\begin{figure}[h]
\begin{center}

\includegraphics[width=3.5in]{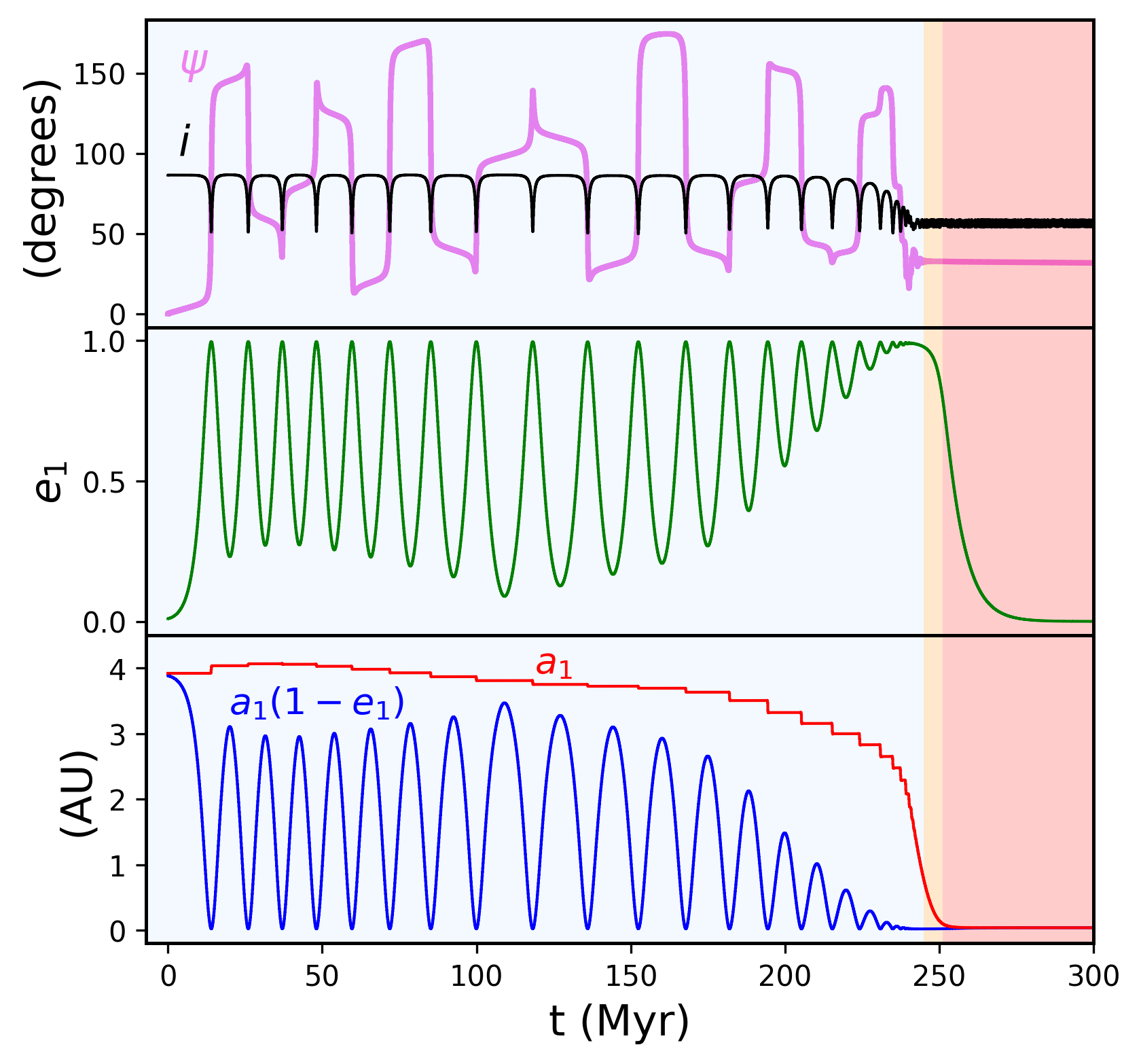}
\caption{\footnotesize Time evolution of the mutual inclination $i$ (black line, upper panel), stellar obliquity $\psi$ (violet line, upper panel), eccentricity $e_1$ (green line, middle panel), semi-major axis $a_1$ (red line, lower panel), and pericenter $a_1(1-e_1)$ (blue line, lower panel) for a simulated initially cold planet that undergoes high-eccentricity tidal migration. We shade the regions where the planet is considered a cold Jupiter (blue region), warm Jupiter (thin orange region), and hot Jupiter (red region). The system is initially set with $m_1 = 1.04 M_{\odot}$, $m_2 = 6.20$$M_J$, $m_3 = 0.94 M_{\odot}$, $a_1 = 3.92$ au, $a_2 = 721.00$ au, $e_1 = 0.01$, $e_2 = 0.55$, $i = 86.4^{\circ}$, $\omega_1 = 82.4^{\circ}$, and $\omega_2 = 141.1^{\circ}$.} 
\label{fig:time_evolution}
\end{center}
\end{figure}

In Fig. \ref{fig:time_evolution}, we show the time evolution for an example system that undergoes high-eccentricity tidal migration in our simulations. The planet starts as an initially cold Jupiter (blue region) in a stellar binary, and its eccentricity and inclination are then excited by EKL. The planet undergoes close pericenter passages, in which tides shrink the semi-major axis. As the semi-major axis becomes smaller, general relativity overwhelms and ultimately quenches the eccentricity and inclination oscillations induced by EKL. The inclination and obliquity become "frozen-in" at this point. Tides then quickly reduce the eccentricity and semi-major axis along a track of nearly constant $a_1(1-e_1^2)$. The planet briefly passes through the warm Jupiter regime (thin orange region) before finally arriving as a close-in hot Jupiter (red region). 

\end{document}